\documentclass{aa}

\usepackage{graphicx}
\usepackage{txfonts}
\usepackage{comment}

\usepackage{xcolor}
\usepackage{csquotes}

\begin{document} 

   \title{Mapping AGN winds: a connection between radio-mode AGN and the AGN feedback cycle}
   \titlerunning{BPT-selected AGN in MaNGA}


   \author{M. Alb\'an \inst{1}
          \and
          D. Wylezalek\inst{1}
          \and J. M. Comerford\inst{2} \and J. E. Greene\inst{3}
          \and R. A. Riffel\inst{4,5}
          }

   \institute{\inst{1}Zentrum f\"ur Astronomie der Universit\"at Heidelberg, Astronomisches Rechen-Institut, M\"onchhofstr, 12-14 69120 Heidelberg, Germany\\
   \inst{2}University of Colorado Boulder, 2000 Colorado Avenue, Boulder, CO 80309, USA\\
   \inst{3}Department of Astrophysical Sciences, Princeton University, 4 Ivy Lane, Princeton, NJ 08544, USA\\
   \inst{4}Departamento de Física, Centro de Ciências Naturais e Exatas, Universidade Federal de Santa Maria, 97105-900, Santa Maria, RS, Brazil\\
   \inst{5}Laboratorio Interinstitucional de e-Astronomia - LIneA, Rua Gal. José Cristino 77, Rio de Janeiro, RJ - 20921-400, Brazil\\
             }

   \date{Received Month xx, 202x; Month Xxxxxx xx, 202x}

  \abstract{We present a kinematic analysis based on the large Integral Field Spectroscopy (IFS) dataset of SDSS-IV MaNGA (Sloan Digital Sky Survey / Mapping Nearby Galaxies at Apache Point Observatory; $\sim$ 10.000 galaxies). We have compiled a diverse sample of 594 unique Active Galactic Nuclei (AGN), identified through a variety of independent selection techniques, encompassing radio (1.4 GHz) observations, optical emission line diagnostics (BPT), broad Balmer emission lines, mid-infrared colors, and hard X-ray emission. We investigate how ionized gas kinematics behave in these different AGN populations through stacked radial profiles of the [OIII]~5007 emission-line width across each AGN population. We contrast AGN populations against each other (and non-AGN galaxies) by matching samples by stellar mass, [OIII]~5007 luminosity, morphology, and redshift. We find similar kinematics between AGN selected by BPT diagnostics compared to broad-line selected AGN. We also identify a population of non-AGN with similar radial profiles as AGN, indicative of the presence of remnant outflows (or fossil outflows) of a past AGN activity. We find that purely radio-selected AGN display enhanced ionized gas line widths across all radii. This suggests that our radio-selection technique is sensitive to a population where AGN-driven kinematic perturbations have been active for longer durations (potentially due to recurrent activity) than in purely optically selected AGN. This connection between radio activity and extended ionized gas outflow signatures is consistent with recent evidence that suggests radio emission (expected to be diffuse) originated due to shocks from outflows. We conclude that different selection techniques can trace different AGN populations not only in terms of energetics but also in terms of AGN evolutionary stages. Our results are important in the context of AGN duty cycle and highlight integral field unit (IFU) data's potential to deepen our knowledge of AGN and galaxy evolution.

 }


\maketitle
%

\section{Introduction}
    
     Active galactic nuclei (AGN) have become a common element in galaxy evolution studies \citep{Heckman_2014} and a fundamental engine for supermassive black hole (SMBH) growth \citep{Alexander_2012}. Observational studies have suggested the connection between supermassive black holes and their host galaxies, finding significant empirical correlations between them \citep{Kormendy_2013}. Specifically, the mass of the SMBH has been seen to correlate with fundamental galaxy properties such as the bulge luminosity \citep{Ferrarese2000} and the bulge velocity dispersion \citep{Marconi2003}. Further evidence has shown that star formation rate history in galaxies peaks at $z \sim 2$, exactly where the black hole accretion history (related to AGN activity) is at its height \citep{Madau_2014, Aird2015}. This suggests an interaction (and coevolution) between the AGN and the interstellar medium (ISM) of its host galaxy \citep{Fabian_2012, Morganti_2017}, known as AGN feedback. Indeed, the released energy required for such a massive black hole to have grown is comparable to or greater than the binding energy of the host galaxy itself \citep{Silk1998}, placing AGN in the spotlight as relevant for understanding galaxy evolution \citep[see also][]{Hopkins_2006}. 
     
     A common property of galaxies hosting an AGN is the presence of strong winds or outflows \citep[e.g.,][]{Mullaney_2013, Harrison_2014, Cheung_2016, Wylezalek2020} in the ionized gas. Such outflows can be deployed in the form of collimated jets \citep{Worrall_2006} or as radiatively-driven winds \citep{Netzer_2006} where gas can be ejected and transferred into the host galaxy \citep[see][for a review]{King_2015}. This ubiquitous characteristic is a popular mechanism to explain how AGN feedback works and has been a key parameter introduced to solve theoretical problems faced in cosmological simulations \citep{Somerville_2015, Naab_2017}. For example, one notable application is helping to explain the regulation of star formation in massive galaxies \citep[see also][]{Harrison_2017}. These phenomena (winds or outflows) have been observed in multiple gas phases \citep[e.g.,][]{Aalto_2012, Fiore_2017, Herrera_2019, Baron_2021, Riffel_2023}, from extremely broad X-ray outflow features \citep[reaching fractions of the speed of light;][]{Tombesi_2012} to cold-molecular gas winds \citep[e.g.,][]{Cicone_2014}. 
     
     Even when focusing on one specific phase, outflow signatures can turn out to be very complex \citep[e.g.,][]{Zakamska_2016_2}. In the ionized gas (the main subject of this paper), for example, such outflows display non-gravitational winds with velocity dispersion (FWHM~$>500$~km~s$^{-1}$) that cannot be explained by the intrinsic rotation of the host galaxy or its dynamical equilibrium \citep[][]{Karouzos_2016}. Outflows usually appear in the spectra as secondary spectral components that accompany the main spectral lines \citep[e.g.,][]{Heckman_1981, Mullaney_2013}. Therefore, the shape of these spectral lines can acquire complex features that a single Gaussian profile cannot model. Instead, multi-component fitting procedures have been widely used to characterize outflow signatures \citep[e.g.,][]{Forster2014}. A widely used tracer to study these signatures is the [O~III]$\lambda$5007 emission line. This forbidden emission line is restricted to low-density environments (such as the narrow line region) and can be produced as a result of shocks or photoionization \citep[][]{osterbrock_1989}. 
   
     Much of what has been learned from AGN has been through the study of their ionized gas kinematics. For example, in a large sample of optically selected Type-II AGN, \citet{Woo_2016} found that the velocity dispersion of the outflow as well as the fraction of emission-line ([O~III]) shapes exhibiting multiple components both tend to escalate with an increase in [O~III] luminosity ($L_{\rm [O~III]}$). This is relevant because the $L_{\rm [O~III]}$ has been shown to be a good indicator of an AGN's bolometric luminosity \citep[L$_{bol}$;][]{Heckman_2004, LaMassa_2010}, which is an important parameter to understand the involved energy injection of the AGN's supermassive black hole \citep{Heckman_2014} to the host galaxy. For example, \citet{Fiore_2017} found that the wind mass outflow rate correlates with L$_{bol}$.
     
     Ionized outflows have been routinely found in AGN selected from Infrared \citep[e.g.,][]{DiPompeo_2018}, X-ray \citep[e.g.,][]{Rojas_2019} and optical surveys \citep[e.g.,][]{Wylezalek2020}, to mention a few examples. However, none of the multiple AGN selection techniques today offer an ultimate clean AGN population \citep{Padovani_2017_at_all_wave} by itself. Attempts to create more complete AGN samples have shown that different selection techniques can find AGN candidates that other single selection techniques would miss \citep[e.g.,][]{Alberts_2020}. Statistical analysis of AGN selected based on techniques that are limited to a certain wavelength window can suffer from important biases such as obscuration or data coverage. This is not a simple task and different selection techniques (using various wavelengths) find different AGN populations, even with contrasting host-galaxy properties \citep[e.g.,][]{Hickox_2009,Comerford2020,Ji_2022}.
     
      Consequently, the estimated outflow properties and, therefore, AGN feedback studies can be compromised by the way the AGN population is selected. For example, \citet{Mullaney_2013} found that the most extreme [O~III] kinematics arise from AGN with moderate radio luminosities ($10^{23}$~W~Hz$^{-1}>$~L$_{1.4~ \text{GHz}}$~$>10^{25}$~W~Hz$^{-1}$), finding evidence of compact radio cores being responsible for driving the most broadened profiles \citep[see also][]{Jarvis_2019, Molyneux_2019, Jarvis_2021}. \citet{Baron_2019} found that AGN that present outflows (using the [O~III] emission line) exhibit an excess in the mid-infrared spectral energy distribution component, suggesting that outflows are carrying dust. Different selection techniques can also be sensitive to different AGN powering mechanisms or stages of the current AGN duty cycle. The latter has been suggested by comparing directly between optically selected AGN candidates against mid-infrared radio-detected AGN candidates \citep[see][]{Kauffmann_2018}, finding the former to dominate black hole growth in lower mass systems.

     An additional complication is that ionized outflows can extend from sub-kpc \citep[e.g.][]{Singha_2022} to kpc scales \citep[e.g.,][]{Liu_2010,Sun_2017}. Due to the limitations of the instruments, most of the studies mentioned above base their results on single-fiber observations. Integral field spectroscopy (IFS) provides a valuable technique to study the spatial distribution of outflows in more detail \citep[e.g.,][]{2018Wylezalek, Luo_2021, Singha_2022}. One of the latest pioneering IFS surveys is the MaNGA (Mapping Nearby Galaxies at Apache Point Observatory) survey \citep{bundy2015}, providing 10~010 unique galaxies with spatially resolved spectra. Hence, our primary objective is to investigate how outflow properties vary, not only spatially but also based on the selection technique employed. The responsiveness of our selection methods to outflow characteristics can potentially shed light on their driving mechanisms and a connection to the AGN duty cycle.     
     
     This paper is organized as follows. In Section \ref{sec_2}, we describe our data and some available catalogs about them that are relevant to this study to assemble a multi-wavelength AGN catalog. The methods employed to study our sample are described in Section \ref{sec_analysis}, with a description of the host galaxy properties of our sample. The results are explained in Section \ref{sec_W80_differences}, and we present a discussion in Section \ref{sec_discussion}. Lastly, we summarize our conclusions in Section \ref{sec_summary}. The cosmological assumptions used in this study are H$_{0}=$72 km s$^{-1}$ Mpc$^{-1}$, $\Omega_{\rm{M}}$=0.3 and $\Omega_{\Lambda}=0.7$.

\section{Sample and catalogs}
\label{sec_2}

\subsection{The MaNGA Survey}
\label{survey_description}

    In this study, we use the $\sim$~10,000 galaxies ($0.01 < z < 0.15 $) observed in the SDSS-IV/MaNGA survey (Sloan Digital Sky Survey / Mapping Nearby Galaxies at Apache Point Observatory). MaNGA is an integral field unit (IFU) survey, providing 2D mapping of optical spectra at 3622--10354 \r{AA} at a resolution of  R$\sim$2000. It's field-of-view ranges from 12\arcsec to 32\arcsec in diameter. Data reduction has been performed by MaNGA's Data Reduction Pipeline \citep[DRP,][]{Law2015}. Complete spectral fitting is provided by MaNGA's Data Analysis Pipeline \citep[DAP,][]{MangaDAP}. The DAP fits models for multiple spectral components (e.g., stellar continuum, emission lines) to the entire spectra. Throughout this paper, we are using the spectra (reduced by the DRP) after subtracting their stellar continuum (i.e., emission line-only spectra provided by the DAP; see details in Section \ref{sec_analysis}).
    
    Additionally, \citet{Sanchez_2022} presents a comprehensive catalog reporting multiple characteristics and integrated host galaxy properties based on a full spectral analysis with the \texttt{pyPipe3D} pipeline \citep{Lacerda_2022}. Most of the galaxy properties used in our study are taken from this catalog (e.g. stellar mass, star formation rates). Other galaxy properties, such as emission-line ratios, H$\alpha$ equivalent widths (EW(H$\alpha$)), are taken from \citep{Alban2023}. In this paper, we furthermore compute additional parameters as described in Section \ref{sec_analysis} (e.g., L$_{[O~III]}$).

\subsection{AGN catalogs}
\label{AGN_catalogs}
    This paper aims to assess the behavior of spatially resolved ionized gas kinematics in AGN samples selected through various selection methods \footnote{Throughout this paper, we will refer to ``AGN population" as a group of AGN chosen by a specific observational technique rather than speaking about a particular type, mode or class (except we state the opposite).}. We use the following set of MaNGA-AGN catalogs that we further describe in the following subsections:
\begin{itemize}
    \item An optical emission line-based catalog from \citet{Alban2023} (using the 2~kpc aperture).
    
    \item A broad-line based AGN catalog from \citet{Fu_2023}.
    
    \item A mid-infrared selected AGN catalog from \citet{Comerford_2024}.
    
    \item A hard X-ray-selected AGN catalog from \citet{Comerford_2024}.
    \item A catalog of radio-selected AGN that we construct in this paper (see section \ref{sec_radio_selection}). 
\end{itemize}

    The full MaNGA sample contains a small number of repeated observations, most of which can be identified through their MaNGA-IDs (although there are exceptions; see more in Appendix \ref{appendix_repeated_targets}). We exclude duplicate sources in our final statistics, tables, and figures. In the following sections, we describe the individual AGN catalogs and the respective selection criteria in more detail. The sky coverage of the different surveys used for the classifications described below overlaps with MaNGA.

\subsubsection{DR17 optical AGN catalog in flexible apertures from Alb\'an \& Wylezalek} 
\label{optical_class}

    \cite{Alban2023} present galaxy classifications based on optical emission line diagnostics \citep{BPT, Veilleux1987} measured within apertures of varying size for the entire MaNGA survey. Galaxies are classified into Star-forming (SF), Composite, Seyfert, LINER \citep[low-ionization emission-line region,][]{Halpern1983}, or Ambiguous galaxies (e.g., if a galaxy received two different classifications based on different line ratio diagnostics). The final AGN sample is then defined based on the galaxies in the Seyfert and LINER classes with an additional cut on H$\alpha$ equivalent width $>$3~\r{AA}\footnote{The equivalent width of each galaxy is obtained using the same aperture size used to measure the emission-line ratios during the classification. Equivalent widths and emission-line ratios are also included in \citet{Alban2023} catalog.}. This additional cut minimized the contamination of faint `fake' AGN \citep{fernandes2010}.

    In this paper, we use the 399 AGN candidates from the catalog based on a 2~kpc aperture\footnote{Note that the catalog from \citet{Alban2023} originally reports 419 targets using a  2~kpc aperture. However, we excluded galaxies due to duplication or critical flags (see Appendix \ref{appendix_repeated_targets}).}. The aperture is chosen to keep a balance between MaNGA's spatial resolution limit \citep[$\sim1.37$~kpc,][]{Wake2017} and the physical extent of gas ionized by an AGN \citep[known as the narrow line region (NLR),][]{Bennert2006, Netzer2015}.
    
\subsubsection{Broad-line AGN catalog}
\label{sec_new_broad}

    Some active galaxies present broad Balmer emission lines \citep[known as Type-I AGN; e.g.,][]{Oh_2015}. This is attributed to Doppler broadening due to high-velocity ionized gas surrounding the SMBH \citep{Peterson_2006}. \citet{Comerford2020} presented a crossmatch between the MaNGA survey and \citet{Oh_2015}'s Type I classification which is based on SDSS DR7 data single-fiber spectroscopic observations with a size of 3''. More recently, \citet{Fu_2023} has carried out an analysis to identify broad-line AGN and double-peaked emission line signatures for the total MaNGA sample using the DR17 data release. MaNGA not only uses smaller fibers (2'') but also provides additional spatial information.
    
    For each galaxy, \citet{Fu_2023} use DAP flux residuals to compare them to the original flux in specific spectral regions (with a size of 20~\r{A}) corresponding to the location of H$\alpha$ and [O~III] emission lines to assess the quality of the DAP's fitting procedure. They arrange the sample in 20~$S/N$ bins (G-band $S/N$ from the DAP) and select galaxies with residuals $> 1 \sigma$ of the residual distribution at each $S/N$ bin \citep[see details on ][]{Fu_2023}. They then perform a spectral fitting on this sample of 1652 galaxies, allowing multiple components to be fitted to emission lines.
    
    They ultimately select broad-line AGN as galaxies where the emission line width ($\sigma$) of the broad component is at least 600~km~s$^{-1}$ larger than the emission line width of the narrow component and present a catalog of 139 broad-line AGN (Type-I) candidates. We find a few duplicate galaxies in this catalog observations (see Appendix \ref{appendix_repeated_targets}), which reduces the sample to 135 targets. The work by \citet{Fu_2023} almost doubles the number of broad-line selected galaxies presented \citet{Comerford2020}; on the other hand, 21 galaxies presented in \citet{Comerford2020} are not found in \citet{Fu_2023}. Discrepancies in the latter context can be related to the difference in FWHM($H\alpha$) constraints and possibly effects from changing look AGN \citep[see][for a review]{Ricci_2023}.

\subsubsection{Mid-IR and X-ray AGN catalogs of Comerford et al.}
\label{sec_2.1}

\citet{Comerford_2024} cross-match MaNGA galaxies with known AGN candidates from multi-wavelength surveys \citep[as in][]{Comerford2020}. For this study, we will use the following catalogs:

\begin{itemize}
    \item  Mid-Infrared AGN catalog based on observations with Wide-field Infrared Survey Explorer \citep[WISE]{Wright_2010}: 123 AGN.
    \item X-Ray selected catalog based on observations with the Burst Alert Telescope \citep[BAT]{Barthelmy_2005}: 29 AGN.

\end{itemize}

    \citet{Comerford_2024} also provides a radio-AGN catalog and a broad-line (Type-I) AGN catalog, which we choose not to use due to our science goals (see \ref{sec_radio_selection} and Section \ref{sec_new_broad}, respectively). Due to repeated observations or critical flags (from the MaNGA DRP; see the details in Appendix \ref{appendix_repeated_targets}) we exclude 7 galaxies from the Mid-Infrared-selected (3 were repeated) and 1 from the X-ray-selected (1 has a critical flag).

\begin{figure}
	\includegraphics[width=\columnwidth]{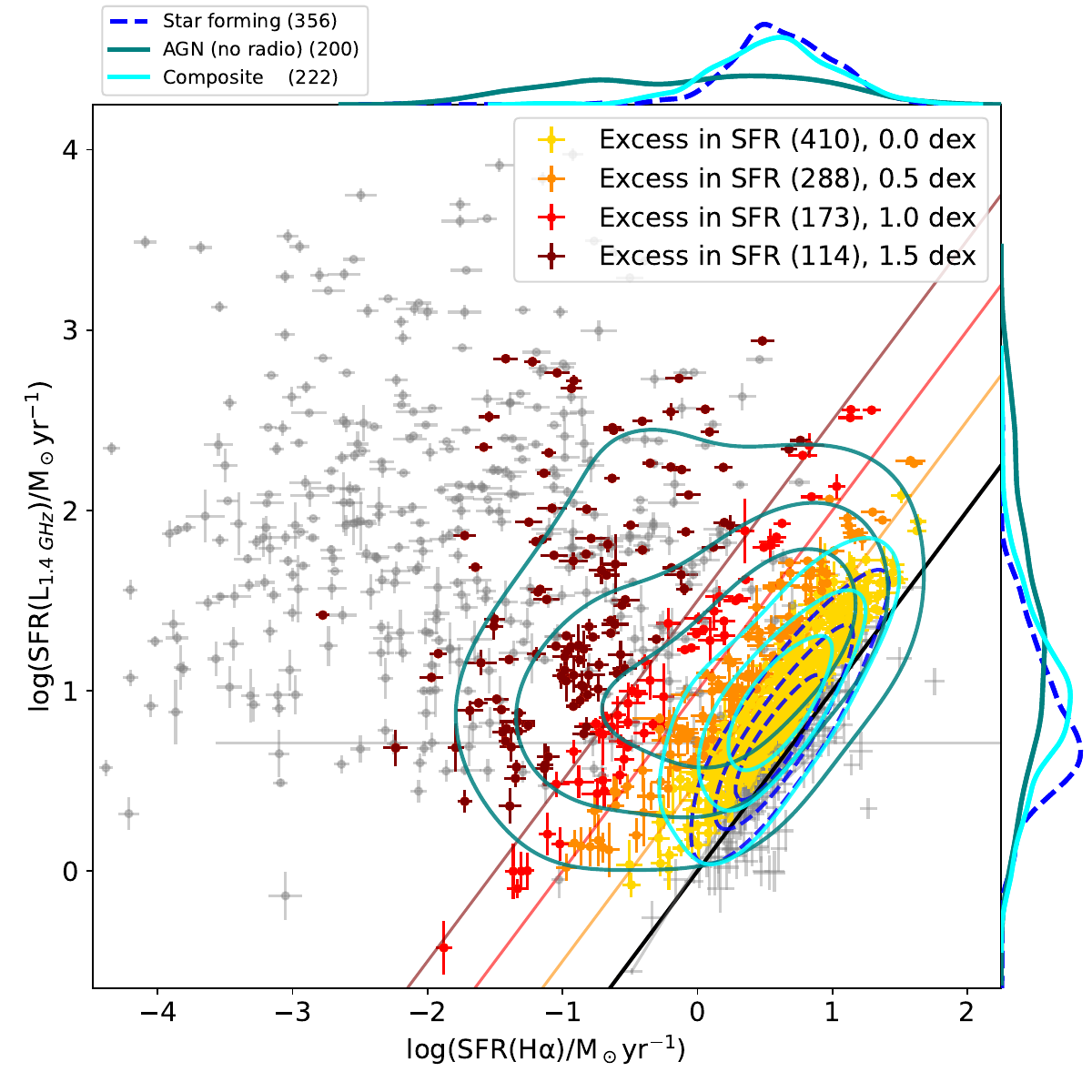}
    \caption{Definition of the radio-selected AGN candidates. We plot in the y-axis the expected SFR that one would measure, assuming that all the radio luminosity can be attributed to star formation processes (SFR(L$_{\text{rad}}$; see Section \ref{sec_radio_selection}). Similarly, in the x-axis, the SFR is expected from H$\alpha$ luminosity (SFR(H$\alpha$)). The black line corresponds to the location where $\text{SFR}(L_{\text{rad}}) = \text{SFR}(\text{H}\alpha)$. The other colored lines (orange, red, and blue) correspond to our SFR excess definition in $\text{SFR}(\text{H}\alpha)$ steps of 0.5, 1.0, and 1.5 dex (log($ x_{i}$)). We define each sample of AGN-selected candidates by selecting the targets whose values (as well as their error bars) are above the corresponding line (following:  $\text{SFR}(L_{\text{rad}})/\text{SFR}(\text{H}\alpha) = x_{i}$) and we color them according to the colored lines (except for the $ x_{i}=0.0$~dex, which corresponds to the yellow ones). We also show targets that did not satisfy any SFR-excess criteria (gray without marker) or did not pass the S/N criteria (gray with marker) for our kinematic analysis (See section \ref{sec_quality_cirteria}). The contours (dashed-blue, light-blue, and teal) represent the density where specific galaxy populations gather (Star-forming, Composite, Non-radio selected AGN). The top and right-hand plots show the individual parameter distribution of these three galaxy populations.}
        \label{fig_radio_sample}
\end{figure}

 \begin{figure}
	\includegraphics[width=\columnwidth]{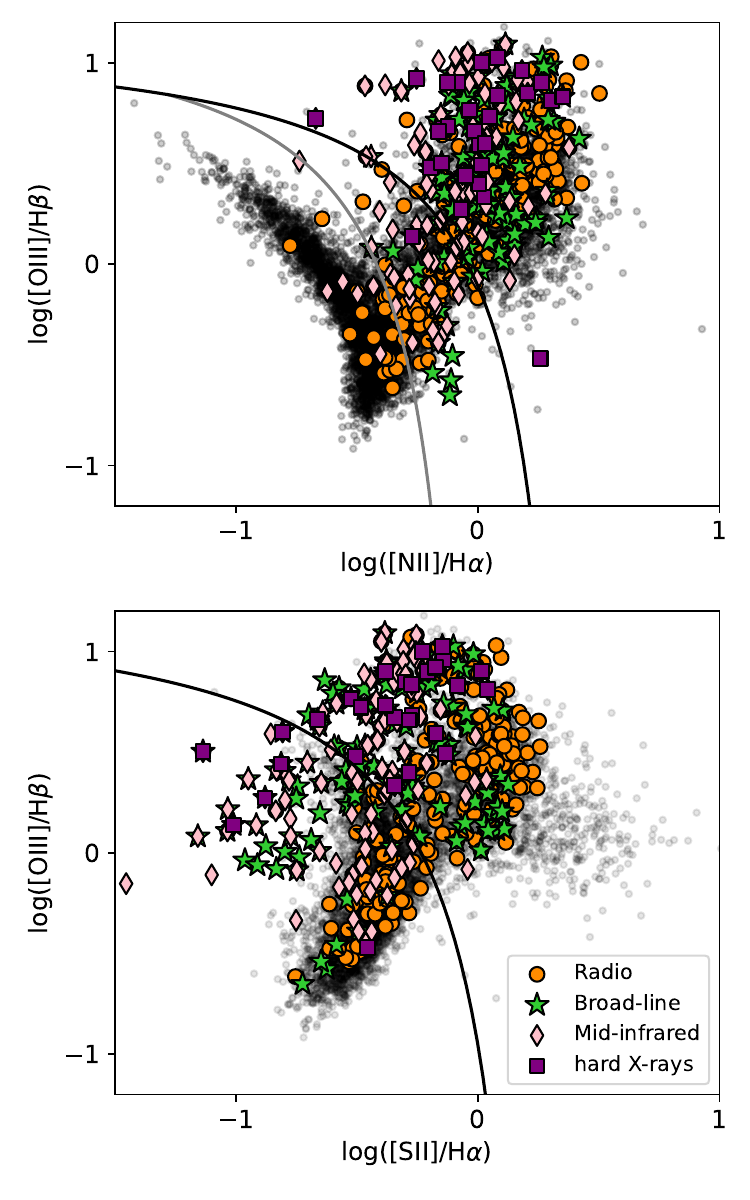}
	
    \caption{Optical diagnostic diagrams for MaNGA galaxies. The black circles in the scatterplot show the emission line ratio values for all MaNGA galaxies. In colored shapes, we feature the different AGN-selected candidates (see the legend). We take the flux ratio values from the ones measured around their central 2~kpc region \citep{Alban2023}. In the top panel, the gray line corresponds to the demarcation line from \citet{Kauffmann2003a}, and the black line (both in the top and bottom plot) corresponds to the ones from \citet{kewley2001}.}\label{fig_BPT-vs-comerford}
\end{figure}

\subsubsection{Selection of radio AGN}
\label{sec_radio_selection}

    In this section, we present a catalog of AGN candidates solely based on radio data and independent of any radio-loud/radio-quiet classification often used in the literature. Broadly speaking, independent of having a jet or not \citep[see the details in some comprehensive reviews; e.g.,][]{Heckman_2014, Padovani_2016A, Panessa_2019}. For example, \cite{Best_2012} used a selection technique combining both optical and radio signatures, and this is the catalog that \cite{Comerford_2024} presents as the radio-selected MaNGA AGN population (see Section \ref{sec_2.1}). However, this classification prioritizes sources with $f_{1.4~\text{GHz}}>5$~mJy as its emphasis lies on the radio-loud population of AGN \citep[see][for a review]{Urry_1995, Padovani_2017}.
    Instead, we use a different approach and first crossmatch MaNGA SDSS-IV galaxies with data from the NRAO Very Large Array Sky Survey \citep[NVSS]{Condon_1998} and the Faint Images of the Radio Sky at Twenty centimeters \citep[FIRST]{Becker_1995} radio surveys and adopt a less strict flux cut of $>1$~mJy. We note that this threshold is close to the sensitivity of the surveys, and there is likely a population of AGN emitting even below this limit \citep[e.g.,][]{White_2015}.
    
    Previous studies have shown that finding genuine associations between targets of two different surveys comes with a trade-off between completeness and reliability \citep[e.g.,][]{Best_2005, Ivezic_2002}. Choosing a larger offset for searching counterparts can lead to high completeness but increases the number of false associations. To ensure precise spatial alignment between the optical and radio sources, we search for the closest galaxy within a 1.5~arcsecond aperture for the FIRST survey, which ensures 85\% of completeness and 97\% reliability \citep[see][]{Ivezic_2002}, and we use a 5.0~arcseconds aperture for the NVSS survey. For comparison, \citet{Best_2005} estimates 90\% of completeness and 6\% of contamination from random targets using a 10.0~arcseconds aperture. This choice of apertures aims to maximize completeness and reliability. We find 936 and 1035 cross-matches for the FIRST and NVSS survey, respectively. In total, 1383 unique targets were found, with 588 coincident targets between FIRST and NVSS.

     Our aim is to develop a radio-selected sample as independent as possible of known optical diagnostics (e.g., BPT diagrams) or other selection criteria. Two significant contributors to the extragalactic radio sky are star formation processes and nuclear activity in galaxies \citep{Padovani_2016A}. AGN host galaxies have been shown to span several decades in radio luminosity, often so faint, raising the question of whether their radio emission is dominated by star formation processes \citep{Panessa_2019} rather than the AGN event. \cite{Zakamska_2016} studied whether the radio signatures (1.4~GHz flux density) of confirmed AGN can be explained purely by star formation processes when comparing it with a variety of star formation rate tracers. Independent of the used SFR tracer, they conclude that AGN had a systematic excess in radio luminosity not consistent with star formation, which might be attributed to the activity in the nucleus. Similarly, \citet{Kauffmann_2008} used the H$\alpha$ luminosity and compared it with the 1.4~GHz flux densities of a sample of SDSS galaxies (cross-matched with radio surveys; FIRST and NVSS) and found that SF galaxies form a tight correlation between both parameters and that their AGN candidates systematically exceed this tight relation towards higher 1.4~GHz flux densities.

\begin{table*}
\centering
	\caption{Coincident targets between the different AGN-selection techniques in the full MaNGA sample.} 
\label{Tablecrossmatch}
\begin{tabular}{c|c|c|c|c|c}
\hline
Selection technique  & Optical & Mid-infrared & hard X-rays & Broad-line & Radio\\
\hline
Optical     & 399 & -   & -  & -   & -    \\
Mid-infrared         & 65  & 123 & -  & -   & -    \\
hard X-rays       & 23  & 24  & 29 & -   & -    \\
Broad-line   & 85  & 48  & 14 & 135 & -    \\
Radio  & 135 & 55  & 17 & 64  & 642  \\
\hline
\end{tabular}
\tablefoot{Here, we have already excluded targets that were repeated observations or targets that had a critical flag (see Appendix \ref{appendix_repeated_targets}).}

\end{table*}

\begin{table*}
    \centering
	\caption{Coincident targets between the different AGN-selection techniques in a sample of MaNGA targets limited by S/N.}
\label{TablecrossmatchSN}
\begin{tabular}{c|c|c|c|c|c}
\hline
Selection technique  & Optical & Mid-infrared & hard X-rays & Broad-line & Radio\\
        \hline
        Optical       & 373 & -   & -  & -   & -   \\
        Mid-infrared  & 64  & 119 & -  & -   & -   \\
        hard X-rays   & 22  & 23  & 27 & -   & -   \\
        Broad-line    & 83  & 48  & 14 & 131 & -   \\
        Radio         & 128 & 55  & 16 & 61  & 288 \\

\hline
\end{tabular}
\tablefoot{Here, we only use targets that satisfy the quality criteria described in Section \ref{sec_quality_cirteria}.}
\end{table*}

     Based on the findings described above, we construct a radio-selected AGN sample similar to the \enquote{L$_{\text{H}\alpha}$ versus L$_{rad}$} method used in \citet{Best_2012}. Specifically, we identify AGN activity based on the excess in the SFR estimated based on the radio luminosity compared to the H$\alpha$-based SFR as reported in the PIPE3D value-added catalog of \cite{Sanchez_2022}; i.e., values that are above the expected 1-to-1 relation. We use the extension named log\_SFR\_SF, meaning that only the spaxels that were consistent with star formation regions were used to measure the SFR. Additional ways to minimize or correct for the contribution of the AGN during the SFR measurement have been shown in De Mellos in prep. We note that different methods to estimate the SFR from optical spectra (e.g., SSP-method, \cite{Sanchez_2022}) do not change our results significantly \citep[as has also been seen in][]{Zakamska_2016}.

    In Figure \ref{fig_radio_sample}, we show the relation between the H$\alpha$-based SFR and the radio-based SFR for the radio-detected MaNGA galaxies described above, assuming all radio emission is related to SF processes. We also show the density contours of different galaxy sub-classes based on optical diagnostics presented in \cite{Alban2023} and described in Section \ref{optical_class}. Pure SF galaxies agglomerate close to the 1-to-1 line (blue dashed contours), in agreement with the findings presented in \citet{Kauffmann_2008}. Composite galaxies (see Section \ref{optical_class}) occupy values consistent with a radio excess (cyan solid contours). Indeed, the emission in composite galaxies is expected to be a mix of star formation and AGN processes. We also show the location of AGN candidates that have been selected by any of the described selection techniques (apart from a radio selection), i.e.,  using mid-infrared, hard X-rays (see Section \ref{sec_2.1}), broad lines (see Section \ref{sec_new_broad}) and optical diagnostics (see Section \ref{optical_class}) and label this sample as \enquote{AGN (no radio)} 
    in the figure (teal-colored solid line). We show that this AGN population gathers preferentially in the excess region of the plot, consistent with the findings of \cite{Zakamska_2016} and \citet{Kauffmann_2008}. On the top and right-hand borders of the plot, we show the distribution of the individual SFR values using a smooth histogram.
    
    Using offsets from the 1-to-1 line, following $\text{SFR}(L_{\text{rad}})/\text{SFR}(\text{H}\alpha) = x_{i}$, we have colored with yellow, orange, red and maroon the galaxy populations with excesses from $ log(x_{i})= 0.0$ to $ log(x_{i})=1.5$~dex. The gray-colored targets in the plot represent the galaxies that we exclude from our analysis due to one or two of two reasons: they were not above the 1-to-1 relation, or their signal-to-noise (S/N) from the [OIII]~5007 emission line had a bad quality to be accepted for our kinematic analysis (see Section \ref{sec_quality_cirteria}). 
    
    We define our radio-AGN sample using the galaxies whose $L_{\text{rad}}$ plus associated flux uncertainties were at least 0.5~dex above the 1-to-1 SFR relation. We find 642 galaxies that satisfy this criterion, while 28 of those are either duplicate or critical targets (see Appendix \ref{appendix_repeated_targets}). We note that only 5\% of the SF classified galaxies are above the 0.5~dex line. Employing larger cutoffs risks excluding low-luminosity AGNs. Notably, 25\% of targets identified as AGN by alternative methods, i.e., excluding radio observations, were found below the 0.5~dex line. Taking into account additional quality criteria necessary for our emission line analysis (see Section \ref{sec_quality_cirteria}), we will work with a sample of 288 radio-AGN, which we will refer to as the radio-selected AGN sample in the remaining parts of this paper.

    Furthermore, at the moment of writing this paper, \citet{Suresh_2024} studied a radio-AGN sample (selected from MaNGA in a very similar way as in this paper) and their Eddington ratio to estimate their radio activity. They find that the Eddington ratio distribution within their AGN sample exhibits a significant dependency on stellar mass, whereas it shows no correlation with the specific star formation rate (sSFR) of the host galaxies. This led them to conclude that, at a fixed stellar mass, SFRs of host galaxies do not influence the radio-AGN selection.

\subsubsection{Overlap and discrepancy between the AGN catalogs}
\label{overlap_section}

    In Table \ref{Tablecrossmatch}, we compile the number of galaxies identified as AGN using the various selection methods discussed above, noting that some galaxies were selected as AGN by multiple techniques. In total, we identify a sample of 970 galaxies that are classified as AGN by at least one method.
    
    Since the work of this paper focuses on the ionized kinematics as traced by the [O~III]~5007 emission line in AGN, we require additional $S/N$ cuts on emission line fluxes (e.g., $S/N<7$; see the details in \ref{sec_quality_cirteria}), which reduces the sample we continue to work with. Table \ref{TablecrossmatchSN} lists 594 individual AGN candidates (621 if repetitions or critical targets are not taken into account; see Appendix \ref{appendix_repeated_targets}) that will be used in our kinematic analysis, indicating most AGN selections remain largely unaffected by our $S/N$ criteria, except for the radio-selected sample. We further discuss this in Section \ref{sec_host_gal_properties}.

    It is largely known that no selection technique is free from limitations. For example, optical selection techniques are mostly biased towards unobscured AGN. This spectral window is significantly impacted by absorption and scattering due to the presence of dust and gas that can obscure the central regions of an AGN (where most of its energetic input occurs). Some contaminants to optical selection techniques can be associated with galaxies dominated by post-asymptotic giant branch stars \citep[e.g.,][]{Singh2013}. Furthermore, dilution from the host galaxy can also play a role in missing AGN emission. Given that opacity due to dust is less effective at longer wavelengths, Mid-infrared selection techniques are less affected by dust attenuation. Most of its critical contaminants dominate at larger redshifts; with MaNGA we work with sources at $z<0.5$. Finally, radio selection techniques are also less affected by obscuration. However, low-luminosity AGN can be difficult to distinguish from star-forming processes. \citet{Padovani_2017} provides a broad and comprehensive overview of this topic.
    
    In Figure \ref{fig_BPT-vs-comerford}, we display the emission-line ratio diagrams highlighting our AGN samples, except for optically selected sample. AGN do not show any preferred location on the diagrams \citep[or a specific side of the demarcation lines; see][]{kewley2001, Kauffmann2003a}. Using AGN that were classified differently than optical techniques, only 4\% of the SF galaxies are AGN, 14\% in the case of Composite, and 7\% for Ambiguous. This is an excellent example of the well-known problem that using only one single criterion is insufficient to obtain a complete picture of the AGN population.
    
    We note that the hard X-ray-selected AGN candidates are the smallest sample. This is not surprising, as the BAT's integration time is kept short to fulfill its scientific goals \citep{Barthelmy_2005}. It is also the sample with the largest overlap with the other AGN samples; all X-ray-selected AGN are also selected as AGN in at least one other selection technique reported in this paper. Indeed, x-ray emission appears to be universal in AGN and the emission is not significantly contaminated by its host galaxy \citep[see a detailed discussion in][]{Padovani_2017}. Hence, the optical, infrared, broad-line, and radio selection techniques make up the four biggest AGN sub-samples in our study, with the largest number of independently selected candidates. Additionally, we define a sample of non-AGN galaxies that will be used in the discussion section. Our non-AGN sample contains all MaNGA galaxies that were not selected as AGN by any method used in this paper.

\section{Analysis}
\label{sec_analysis}
Hereinafter, when referring to kinematics, we specifically refer to the ionized gas (traced by the [O~III]~5007 emission line).
\subsection{Fitting procedure}
\label{sec_3.1}
Spectra from regions with kinematics dominated by winds can display complex emission line profiles \citep[e.g.,][]{Liu_2013}. This is not taken into account by the DAP emission-line fitting routine. Therefore, we develop a fitting procedure to account for up to two Gaussian components for each emission line. Our fitting method is based on a least-squares Python program using the documentation from Non-Linear Least Squares Minimization \citep[LMFIT,][]{LMFIT} and it follows standard fitting procedure techniques \citep[e.g.,][]{Liu_2013,Wylezalek2020}. In summary, for all spectra in each MaNGA galaxy, our procedure operates in the rest-frame stellar-subtracted region where the [O~III]~5007 emission line is (see the details in Appendix \ref{app_fitting_procedure}). From the maps of the best-fit parameters, we create a non-parametric emission line width W$_{80}$ map. The W$_{80}$ parameter is the most essential value we extract from our fitting procedure and will be the most relevant for the discussion throughout this paper. Whenever we refer to it, we refer to the W$_{80}$ value obtained from [O~III]~5007. 

To study the spatial distribution of this parameter, we construct radial profiles for all galaxies from elliptical annuli in steps of 0.25 effective radius (R$_{eff}$). To obtain the parameters for the elliptical apertures for each galaxy, we use the $b/a$ axis ratio and position angle (PA) from PIPE3D's value-added catalog from \citet{Sanchez_2022}, as well as the effective radius. These parameters are adopted from the NASA-Sloan Atlas catalog \citep{Blanton_2011}. They use the Petrosian system \citep[see][]{Petrosian_1976, Blanton_2001} applied to the SDSS r-band imaging of galaxies using elliptical apertures. Here R$_{eff}$ is defined as the major axis containing 50\% of the flux inside 2 Petrosian radii, and  $b/a$, and PA are obtained from the elliptical aperture \citep[see the details in][]{Wake2017}. To perform a weighted average on each annulus, we capture the fraction of each pixel enclosed by an annulus so that we avoid average properties over a set of discrete pixels and recover a smooth distribution. Specifically, we follow the pixel-weighted average procedure used in \citet{Alban2023} but using ellipses.

 In a sample of $\sim$160~000 normal SDSS (SF BPT selected) galaxies ($z<0.7$, with $8<log(M_{*}/M_{\odot})<11.5$ and $-3<log(SFR/M_{\odot}~yr^{-1})<2$), \citet{Cicone_2016} find that the gas velocity dispersion ($\sigma$) hardly exceeds 150 km~s$^{-1}$. The latter corresponds to a W$_{80}$ of $\sim$ 380~km~s$^{-1}$ ($W_{80} = 2.56 \sigma$). Furthermore, \citet{Gatto_2024} conclude a lower cut, of $\sim$ 315~km~s$^{-1}$ when studying the W$_{80}$ in a control sample of non-AGN (matched to optically-selected AGN in stellar mass, morphology, inclination and redshift). Therefore, W$_{80}$ values greater than this threshold suggest the presence of non-gravitational motion of gas, such as outflows. 

\begin{figure*}
	
      \includegraphics[width=500pt]
	{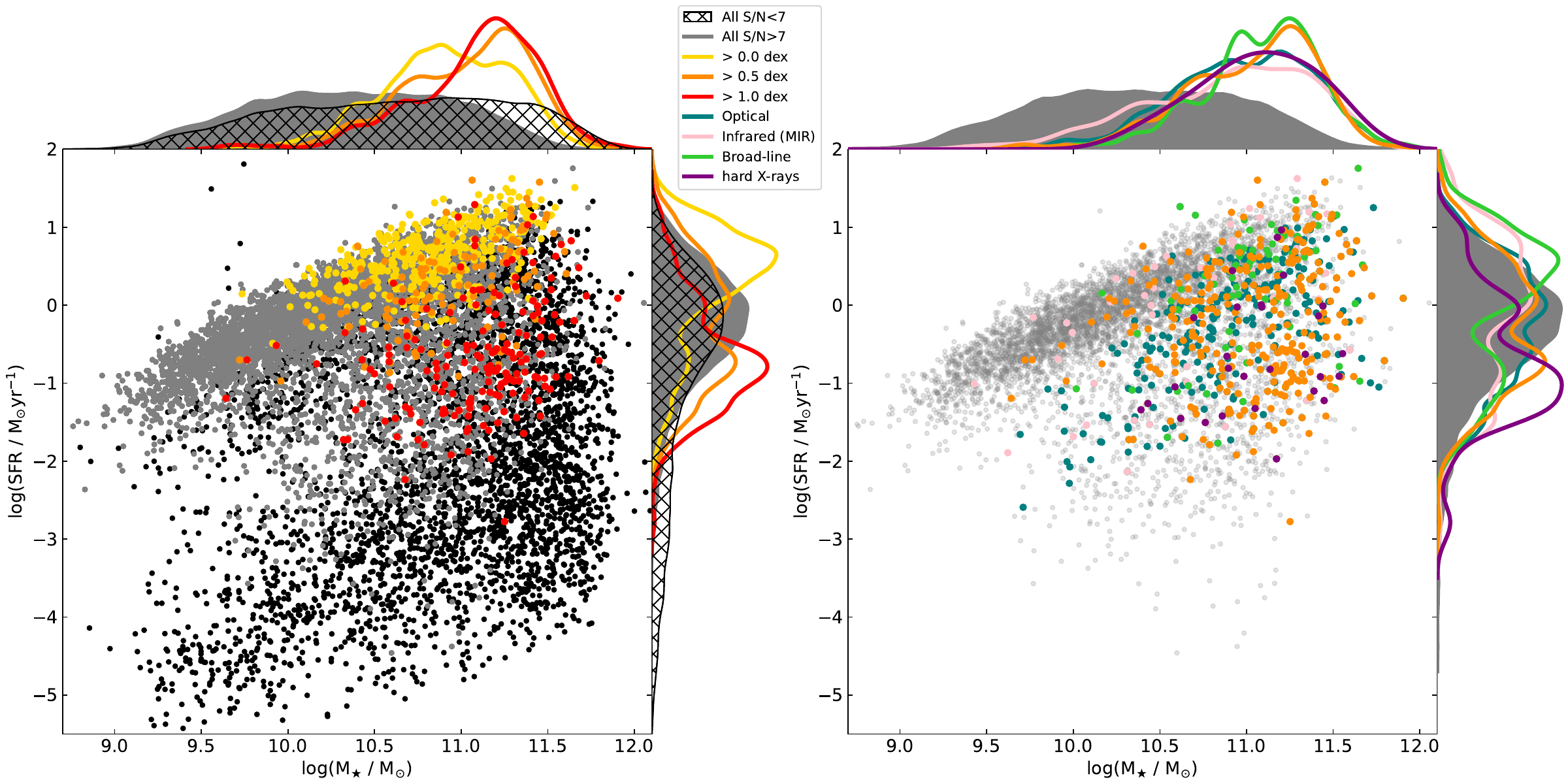}

    \caption{Stellar mass versus star formation rates of MaNGA host galaxies. The left panel shows the impact of our quality criteria (see Section \ref{sec_quality_cirteria}), excluding galaxies with higher stellar mass and lower star formation rates (represented by the black dots in the scatter plot and black hatched distribution in the top and right-hand diagrams). In red, orange, and yellow, we show the distribution of the AGN candidates selected by radio (using 0.0, 0.5, and 1.0 dex of excess; see Section \ref{sec_radio_selection}) that satisfy the $S/N$ quality criteria. To understand the properties of the excluded radio-selected hosts, we encourage the reader to look at Figure \ref{fig_radio_sample}. It can be seen that a long tail of deficient SFR hosts are excluded ($-2>$~log($\text{SFR}(\text{H}\alpha)$)~$>-4$). On the right, we show only the galaxies chosen after the quality criteria and their corresponding AGN classification. We have provided labels for each color in a panel between both plots.}
    \label{fig_MS_s}
\end{figure*}

    \begin{figure*}
	    \includegraphics[width=500pt]
	{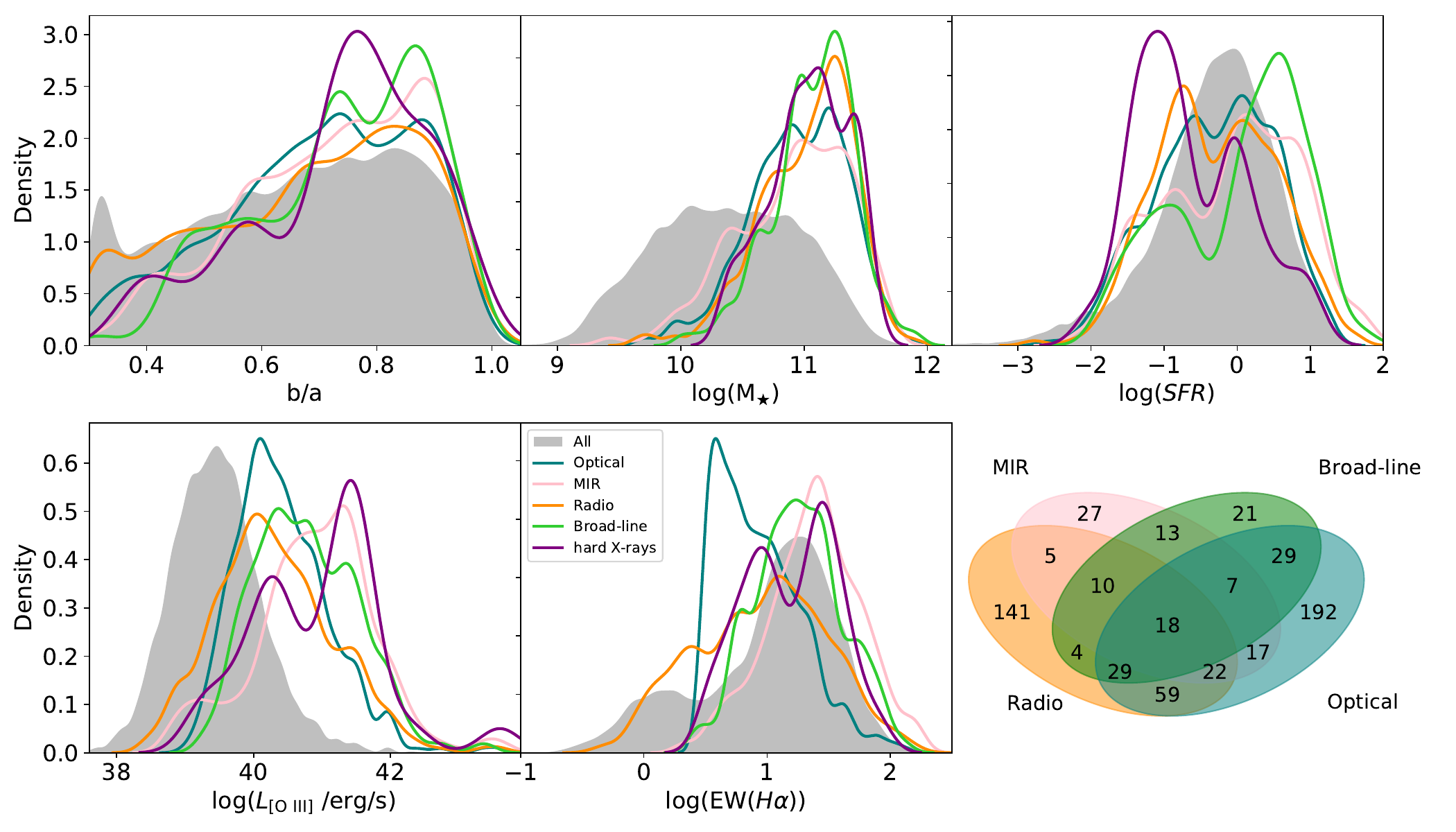}
	
        \caption{Smooth histograms of multiple host galaxy properties. On top (left to right): the $b/a$ axis ratio, log($M_{\star}$), and log(SFR) from Pipe3D. On the bottom (left and middle plot): log($L_{\rm [O~III]}$) and log(EW(H$_{\alpha}$)), both extracted from an aperture of 1~$R_{\rm eff}$, and (to the right) a complementary illustration for Table \ref{TablecrossmatchSN} employing Venn diagrams. In order to maintain visual clarity, the hard X-ray selected AGN sample was intentionally omitted from the diagrams. The gray shaded histogram shows the distribution of all MaNGA that pass our S/N criteria (see Section \ref{sec_quality_cirteria}).}        \label{fig_many_hists}
    \end{figure*}

    \begin{figure*}
	\includegraphics[width=500pt]{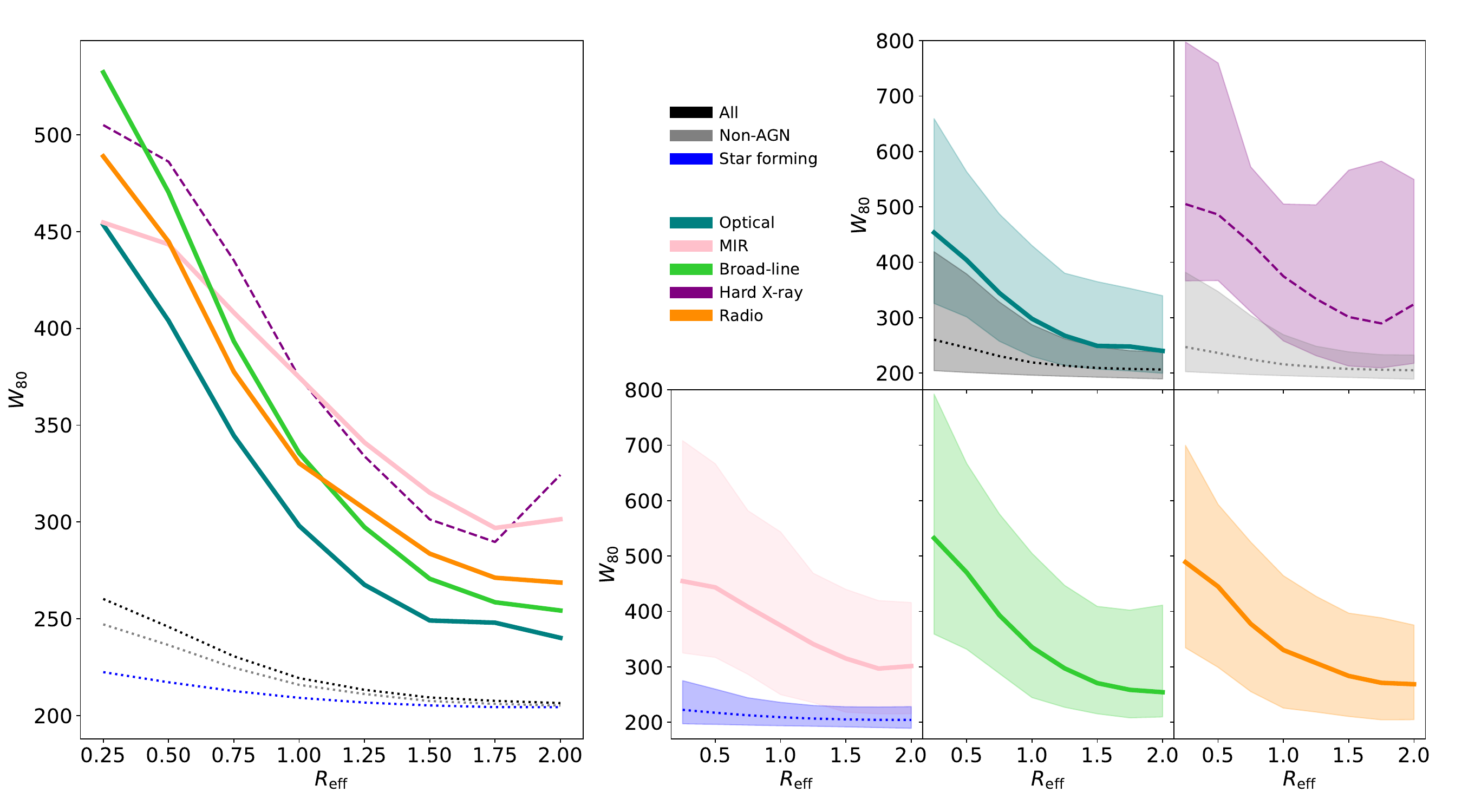}
 
    \caption{W$_{80}$ stacked radial profiles for various galaxy groups. In the upper central panel, colors represent these groups: black, gray, and blue for all MaNGA galaxies, non-AGN, and SF galaxies (selected by BPT), while the remaining colors denote AGN-selected candidates. Each line in the plots represents the median value at each $R_{\rm eff}$ ring, with shaded areas indicating the 14th to 86th percentiles. The leftmost plot displays unshaded profiles for easier comparison. Different line styles are used for visual clarity.}\label{fig_W80s}
    
   \end{figure*}

 \subsection{Galaxies selected for the kinematic analysis based on S/N quality criteria}
 \label{sec_quality_cirteria}

    One crucial factor to consider is the impact of the $S/N$ on measuring the line width W$_{80}$. As $S/N$ decreases, the W$_{80}$ measurements tend to get underestimated \citep[see][]{Liu_2013}, especially if there is indeed a (faint) broad component present in the line profile \citep{Zakamska_2014}. To ensure the accuracy of our analysis and avoid incorrect W$_{80}$ measurements, we exclude all spaxels with an $A/N<7$ (amplitude over noise) before we perform the spectral line fitting. Given the tight relation between S/N and A/N seen in \citet{Belfiore_2019}, we will refer to A/N as simply S/N. To ensure that each individual galaxy retains enough high S/N spaxels, we furthermore apply the following criteria: 

    We only include galaxies in our final sample which satisfy the following criteria:

    \begin{itemize}
        \item More than 10 spaxels with a $S/N>7$.
        \item There are at least two annuli (for the radial profile derivation) where the area covered by the spaxels with a $S/N>7$ is at least 10\% of each annulus's total area.

    \end{itemize}

    The thresholds for the $S/N$ and pixel fractions are chosen to minimize the number of excluded galaxies while retaining a suitable quality for the analysis. These quality cuts introduce a bias (driven by the $S/N$) that rejects more likely low star-forming and some high-mass galaxies, respectively \citep[e.g.,][]{Brinchmann2004, Alban2023}. Our final sample contains 5696 targets (see the left plot on Figure \ref{fig_MS_s}). Table \ref{Tablecrossmatch} and Table \ref{TablecrossmatchSN} show the cross-matches between the different AGN populations and how the subsample sizes decrease after applying the $S/N$ and quality cuts. Radio-selected AGN candidates are significantly impacted by the quality criteria (see Figure \ref{fig_radio_sample}). In contrast, the other  AGN samples remain relatively unaffected.
    
    According to the optical classification from \citet{Alban2023},  more than $90\%$ of the radio-selected-AGN that were excluded from the final sample due to the quality criteria are LINERs ($\sim30\%$, with EW(H$\alpha<3$)) or `lineless' \citep[$\sim60\%$, galaxies that could not be classified by optical analysis, with $S/N<3$; see the details in][]{Alban2023}, and around $5\%$ are classified as ambiguous.

    \begin{figure*}
	    \includegraphics[width=520pt]{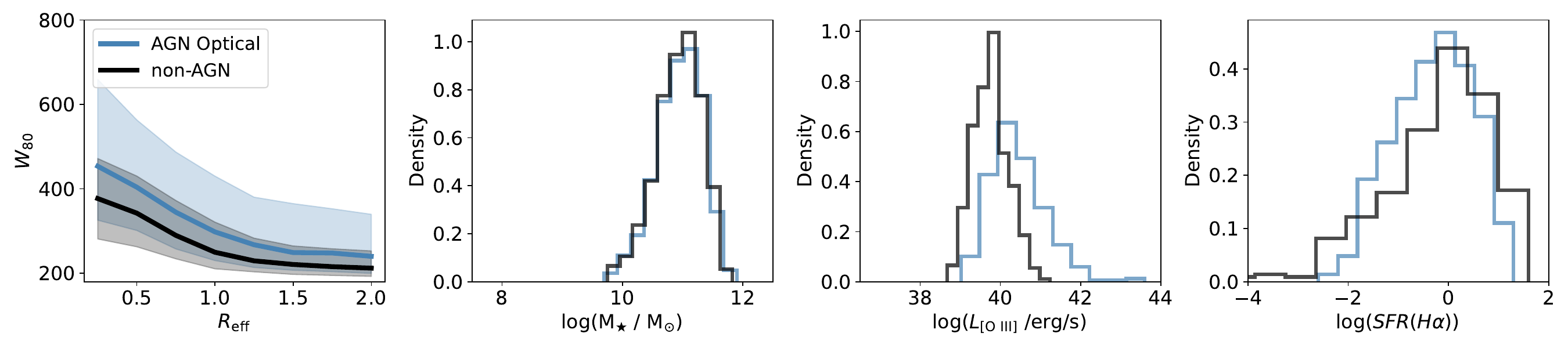}
        \caption{Comparison of stacked W$_{80}$ profiles and host galaxy properties between non-AGN and Optical AGN. In each plot, blue corresponds to the behavior of a specific parameter for Optically selected AGN and black for non-AGN. We show the W$_{80}$ stacked radial profile, the log($M_{\star}$) distribution, the log($L_{\rm [O~III]}$) distribution, and the log(SFR(H$\alpha$)) distribution. For this comparison, only log($M_{\star}$), redshift, and morphology were controlled.}
        \label{fig_nonAGN_vs_Optical_AGN_only_MASS}
    \end{figure*}

\subsection{Typical properties of AGN-selected host galaxies}
\label{sec_host_gal_properties}

    Focusing on the final sample of galaxies that fulfill the $S/N$ and quality criteria, Figure \ref{fig_many_hists} shows host galaxy properties for our AGN populations, including a Venn\footnote{We create our Venn diagrams from an adapted version of this public repository: \url{https://github.com/tctianchi/pyvenn.}} diagram (following to Table \ref{TablecrossmatchSN}). Comparing Table \ref{Tablecrossmatch} with Table \ref{TablecrossmatchSN} (samples before and after the quality cuts) reveals that the AGN samples do not experience a significant cut, with the exception of radio-selected AGN where a notable fraction of massive galaxies, with low SFR and low EW(H$\alpha$) got excluded (see \ref{sec_quality_cirteria}). However, the distribution of $b/a$ and L$_{\text{[O~III]}}$ for radio-selected AGN remains less affected by the S/N cut.

    Most AGN are found in host galaxies with high stellar masses ($M_{\star}$; see the top-middle plot of Figure \ref{fig_many_hists}), regardless of the AGN selection technique. This is a ubiquitous trend that has been found in various AGN samples from different studies \citep[see e.g.][]{Kauffmann2003a, Powell_2018, Barrows_2021, Best_2005}. Our different AGN subsamples all have similar stellar mass distributions. This is an important fact to notice, given that more massive galaxies are expected to have larger emission-line widths \citep[e.g.,][see also Appendix \ref{app_binning}]{Chae_2011, Zahid_2016, Cappellari_2016}. 

    Figure \ref{fig_many_hists} reveals that the different AGN samples probe different distributions of their host SFR, L$_{[O~III]}$, and EW(H$\alpha$), showcasing the biases of each selection technique. The SFR differences between the samples are reflected in Figure \ref{fig_MS_s}, where our AGN candidates tend to gather below the star formation main sequence (SMFS). AGN studies for samples in the local universe have found similar results, where AGN are found in the so-called transition zone or the green valley \citep[e.g.,][]{Schawinski_2007, Salim_2014, Leslie_2016}. At slightly higher redshifts ($0.25 < z < 0.8$), \cite{Hickox_2009} find that mid-infrared selected AGN have bluer colors and are found preferentially in the blue cloud, while radio-selected AGN gather more likely in the red sequence, suggesting the latter are relevant for understanding the evolutionary transition of host galaxies from actively star-forming to more quiescent states.
    
    \citet{2018Wylezalek} find that AGN-selected MaNGA (DR14) targets have mostly low to intermediate luminosities ($L_{[\text{O~III}]}\sim10^{40}$~erg~s$^{\text{-1}}$) for an optically-selected AGN sample. We observe here the same behavior for our optically-selected AGN in MaNGA-DR17. However, for AGN selected via infrared, hard X-rays, or broad Balmer lines, we typically observe higher L$_{[O~III]}$, with distributions peaking at $\sim$~L$_{[\text{O~III}]}\sim10^{41.6}$~erg~s$^{\text{-1}}$. On the other hand, radio-selected AGN candidates show some lower L$_{[\text{O~III}]}$ values. Interestingly, $L_{[\text{O~III}]}$ is known to correlate with the AGN's bolometric luminosity \citep{Heckman_2004, LaMassa_2010, Pennell_2017}, which in turn is correlated with AGN-driven wind velocities \citep[e.g.,][]{Fiore_2017}.
    
   From the 288 radio-selected AGN used for this analysis, 52 of them were optically classified as LINERs with AGN (see Section \ref{optical_class}), and 55 were not classified as AGN as they did not meet the minimum H$\alpha$ equivalent width of 3~\r{AA}. The 52 LINER galaxies have a median EW(H$\alpha$) of $\sim$~1.75~\r{AA}. Other authors have used less strict EW(H$\alpha$) constraints \citep[e.g., 1.5~\r{AA}]{Sanchez_2018} to include fainter AGN in optically selected samples. However, this might introduce some LINER-like galaxies with no AGN but dominated by a population of post-AGB stars \citep{Singh2013} that can mimic AGN-like ionization in typical optical classification. This emphasizes the importance of a multi-wavelength AGN selection technique for a more complete population census. Quite remarkably, AGN that were selected by MIR, broad-lines, or X-ray observations show clearly EW(H$\alpha)>3.0$~\r{A} (only three galaxies have EW(H$\alpha)<3.0$~\r{A} $\sim$  2.0~\r{A}), while radio-selected AGN show lower EW(H$\alpha)\sim$~2.0\r{A}. We recall that optically-selected AGN were required to have EW(H$\alpha)>3.0$~\r{A} \citep{Alban2023}.

    Therefore, we will investigate whether the differences in the W$_{80}$ radial profiles (see Section \ref{w80results}) can be attributed to any differences in host-galaxy properties.

\subsection{Radial profiles of ionized gas kinematics}
\label{w80results}

    Several studies have investigated the overall AGN kinematic properties across many different AGN samples \citep[e.g.,][among others]{Mullaney_2013, Zakamska_2014, Baron_2019, Rojas_2019}. However, most of these studies have used single-fiber data and have, therefore, been limited in assessing spatial dependencies. Consequently, comprehensive studies with large spatially resolved spectral samples are crucial to assessing the impact of selection techniques. We investigate the radial profiles of ionized gas kinematics in the various AGN samples we have defined above. To do so, we first stack the W$_{80}$ profiles (see Section \ref{sec_3.1}) of galaxies within each individual subsample and use the median value at each annulus. 
    
    The resulting profiles are shown in Figure \ref{fig_W80s}. The shaded regions around the profiles represent the 14~$_{th}$ and 86~$_{th}$ percentiles of the W$_{80}$ distribution at each specific annulus. The median W$_{80}$ profiles reveal distinct behaviors between the AGN samples. Visual inspection indicates variations not only in the magnitude but also in the slope of these profiles. Notably, regardless of AGN selection technique, there is a systematic behavior of an enhanced W$_{80}$ profile in the AGN population compared to the overall MaNGA sample. This characteristic continues out to 2 effective radii. Similarly, if we focus only on galaxies that were not selected as AGN by any of our selection techniques (the non-AGN, see Section \ref{overlap_section}), we find the same trend. Moreover, SF-classified galaxies exhibit less pronounced profiles, with minimal enhancements near the center. The subsequent section will explore potential explanations for these observations.


    \begin{figure*}
	    \includegraphics[width=520pt]{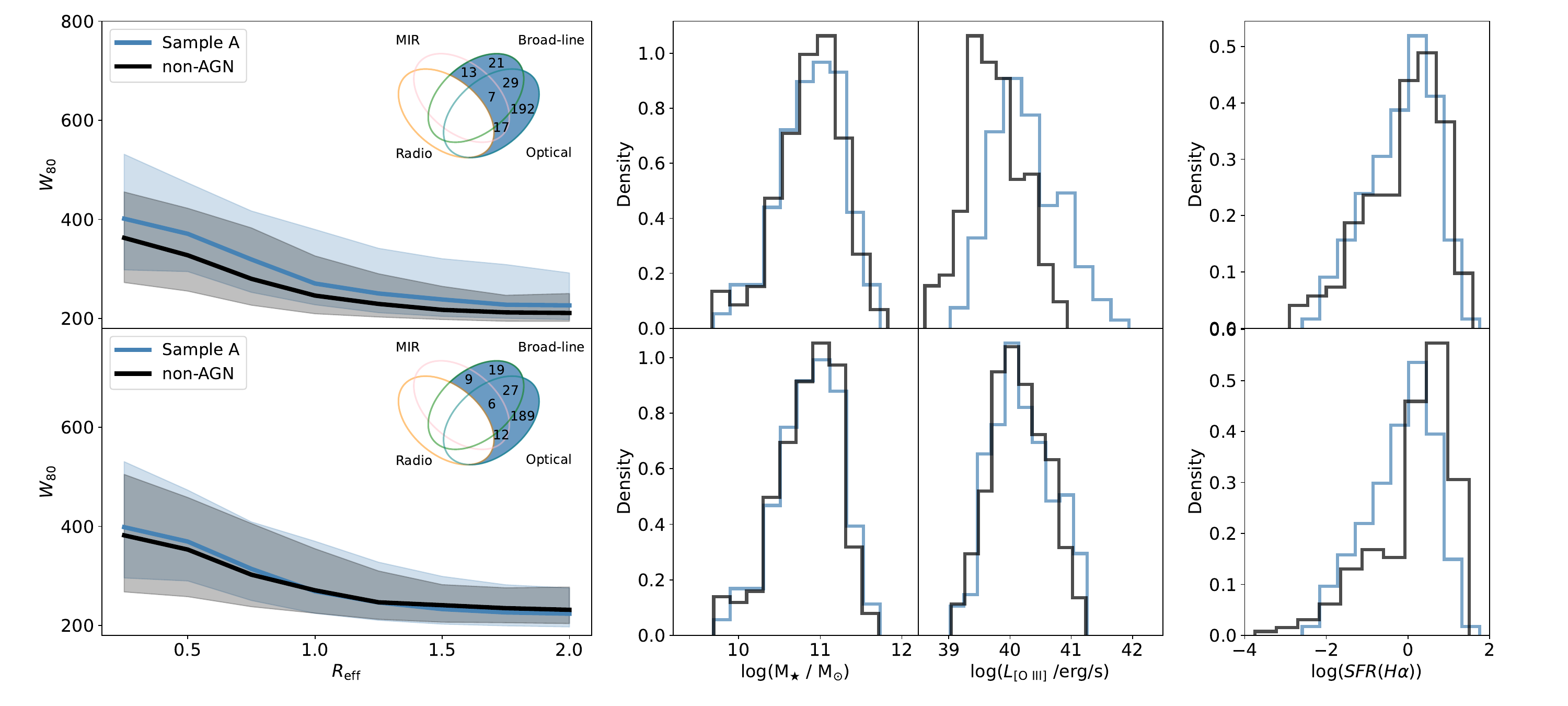}
        \caption{Comparison of stacked W$_{80}$ profiles and host galaxy properties between non-AGN and Sample A. In each plot, blue corresponds to the behavior of a specific parameter for Sample A (see Section \ref{sec:AGN_vs_non-AGN}) and black for non-AGN. For each column of plots, from left to right, we show the W$_{80}$ stacked radial profile, the log($M_{\star}$) distribution, the log($L_{\rm [O~III]}$) distribution, and the log(SFR(H$\alpha$)) distribution. The plots in the bottom row show how both samples behave after they are matched to have the same log($M_{\star}$) and log($L_{\rm [O~III]}$). The plots in the top row show how both samples behave if the matched is done only for the log($M_{\star}$). The Venn diagrams shown in each W$_{80}$ plot represent Sample A. The discrepancy in the Venn diagram numbers in Sample A arises from the incorporation of $L_{\rm [O~III]}$ into the control.}
        \label{fig_SampleA_vs_nonAGN}
    \end{figure*}

    \begin{figure*}
	    \includegraphics[width=520pt]{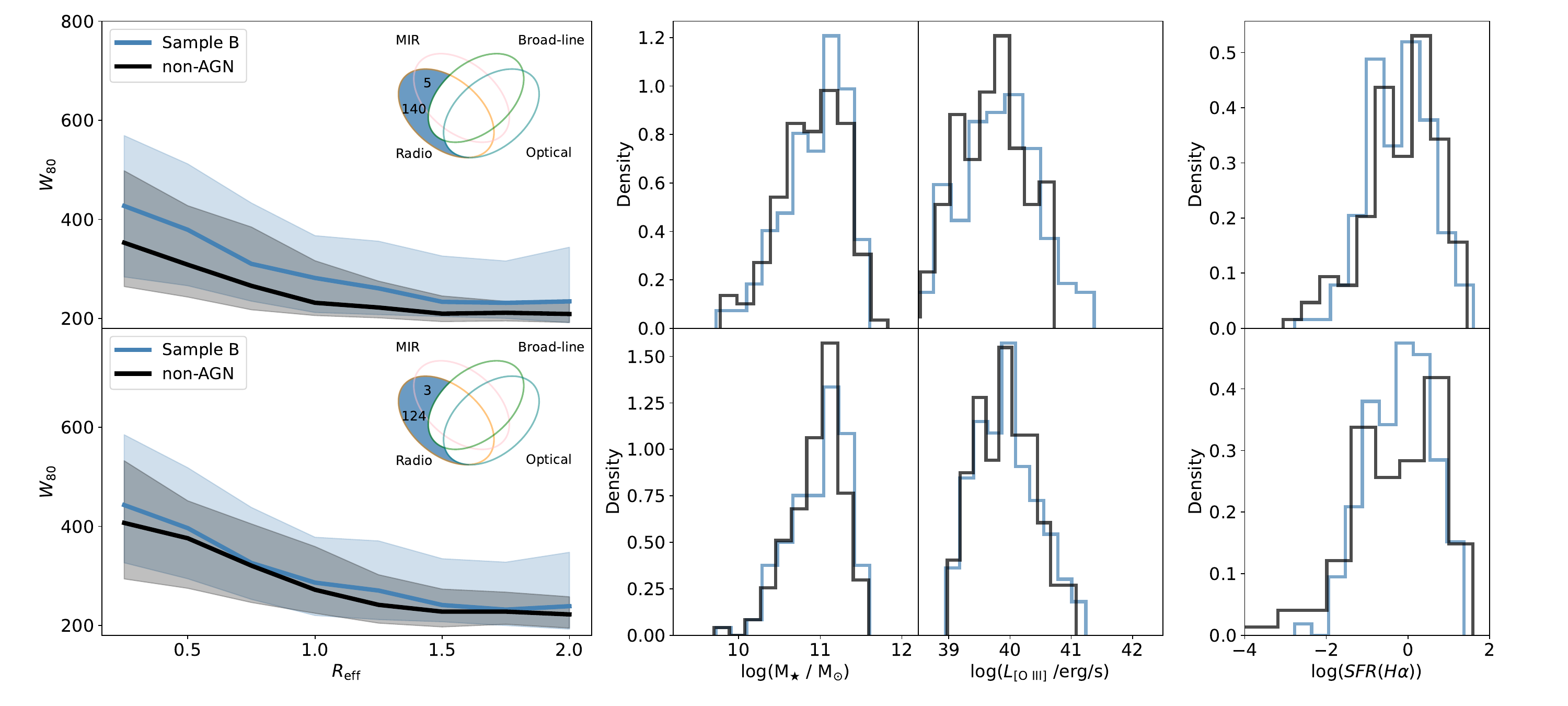}%
        \caption{Comparison of stacked W$_{80}$ profiles and host galaxy properties between non-AGN and Sample B. Same as Figure \ref{fig_SampleA_vs_nonAGN} but comparing non-AGN (black) to Sample~B (blue).}
        \label{fig_SampleB_vs_nonAGN}
    \end{figure*}

\section{What drives the differences in the observed kinematics?}
\label{sec_W80_differences}

        Figure \ref{fig_W80s} (see the left panel) reveals that the median W$_{80}$ radial profiles of AGN-selected populations are significantly different. In Section  \ref{sec_host_gal_properties}, we have shown that the host galaxies of the different AGN samples are similar with respect to some properties (e.g., stellar mass, or $b/a$ axis-ratio) but significantly different with respect to other properties (e.g., $L_{\rm [O~III]}$, or SFR(H$\alpha$)). We note that samples with higher $M_{\star}$ will systematically select galaxies with higher $L_{\rm [O~III]}$ and vice versa (see also Appendix \ref{app_binning}). At the same time, samples with higher $M_{\star}$ and $L_{\rm [O~III]}$ will systematically select galaxies with higher W$_{80}$ (see discussion in Section \ref{sec_host_gal_properties}). Therefore, in this Section, we investigate if and how the differences in the kinematics persist or change when we carefully match the AGN samples so that they have the same host galaxy properties.
        
        We create control samples based on a $M_{\star}$ and $L_{\rm [O~III]}$ parameter space. Given that the number of galaxies per each $M_{\star}$ and $L_{\rm [O~III]}$ bin becomes limited, controlling for redshift and morphology becomes challenging. Therefore, we select the galaxy that is closest in redshift and in morphology. The morphology is used as a number \citep[obtained from][]{Sanchez_2022}. We also note that the radial profiles take the R$_{eff}$ of each galaxy into account by using it as a step for the average W$_{80}$ at each annulus (see Section \ref{w80results}).

    \begin{figure*}
        \includegraphics[width=520pt]{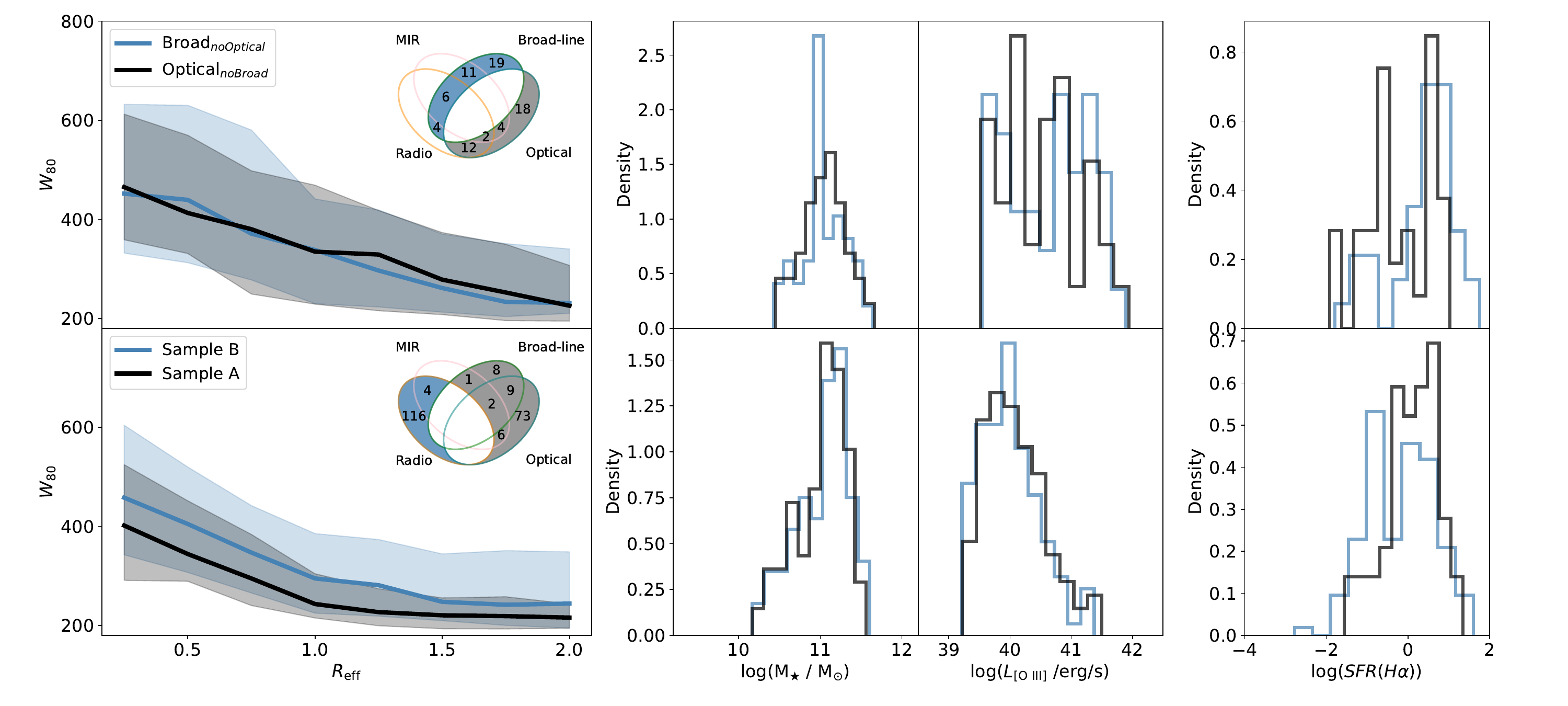}
 
    \caption{Comparison of stacked W$_{80}$ profiles and host galaxy properties between different AGN samples. The plots in the top and bottom rows show how both samples behave after they are matched to have the same log($M_{\star}$) and log($L_{\rm [O~III]}$). The Venn diagrams shown in each W$_{80}$ plot represent the AGN sample in the label. On top: blue for optically-selected (without broad-line-selected) and gray for the opposite. On the bottom: blue for Sample~B and gray for Sample~A. For each column of plots, from left to right, we show the W$_{80}$ stacked radial profile, the log($M_{\star}$) distribution, the log($L_{\rm [O~III]}$) distribution, and the log(SFR(H$\alpha$)) distribution.}\label{fig:agn_vs_agn}

   \end{figure*}

\subsection{AGN versus non-AGN}
\label{sec:AGN_vs_non-AGN}

    In a recent study, \citet{Gatto_2024} used a catalog of optically-selected AGN (selected through emission-line diagnostics) and created a control sample matching properties to the AGN hosts similarly as in this paper, except for the $L_{\rm [O~III]}$. When looking at all spaxels, they find that AGN have greater $W_{80}$ values than the control galaxies, attributing the ionized gas kinematic disturbances to the presence of the AGN. We obtain similar results as in \citet{Gatto_2024} (see Figure 1) when using only optically selected AGN and a similar control sample. In Figure \ref{fig_nonAGN_vs_Optical_AGN_only_MASS}, we present the results as radial profiles for this comparison\footnote{We use our non-AGN sample, i.e., removing also AGN selected by a multi-wavelength selection as discussed in Section \ref{AGN_catalogs}.}, finding that optically selected AGN have larger velocity widths than non-AGN of similar masses. We find similar results when selecting AGN through other techniques (see below).

    \citet{Gatto_2024} find that both, AGN and control galaxies, have $L_{\rm [O~III]}$ values that correlate positively with their average $W_{80}$. We see that their control sample has a median $L_{\rm [O~III]}$ that is $\sim$1~dex lower than the ones from AGN. Therefore, in our analysis, we ask whether the kinematic differences are still present between non-AGN and AGN if taking the $L_{\rm [O~III]}$ into account during the control (this removes the most luminous AGN). Under these conditions, we observe that AGN and controls are more alike in terms of their $W_{80}$ values (see the discussion in Section \ref{sec_discussion}).

    With this approach, we aim to assess the question: is the current nuclear activity responsible for the enhanced radial profiles in the AGN populations? And if so, to what extent are these perturbations spreading? We also seek to investigate how the kinematics may be dependent on the AGN selection technique. Therefore, we look at the $W_{80}$ radial profiles comparing non-AGN (see Figure \ref{fig_SampleA_vs_nonAGN} and Figure \ref{fig_SampleB_vs_nonAGN}) with the following samples:

    \begin{itemize}
        \item Sample A: targets selected via optical or broad lines, excluding the ones selected via radio.
        \item Sample B: targets selected via radio but not via optical or broad lines.
    \end{itemize}

    We do not observe a significant difference (in terms of ionized gas kinematics) between optically selected and broad-line selected AGN (see the details in Section \ref{sec:AGN_vs_AGN}). This is the main reason why we merge them when defining Sample A and Sample B for the purpose of comparing them with radio-selected AGN.

    We present a comparison between non-AGN and AGN, using as a control sample non-AGN first matched only in stellar mass, morphology, and redshift, and, later, including $L_{\rm [O~III]}$ in the parameter match (as mentioned at the beginning of this section). Figure \ref{fig_SampleA_vs_nonAGN} shows two rows of plots. The top row shows the W$_{80}$ radial profiles as well as histograms of host galaxy properties of Sample~A and non-AGN without including $L_{\rm [O~III]}$ during the match. The bottom row shows the comparison considering the $L_{\rm [O~III]}$ during the match. Note that the latter matching procedure leads to the removal of the most extreme AGN in Sample~A exhibiting the highest  $L_{\rm [O~III]}$. The shaded areas in the radial profiles show the 14th and 86th percentiles as in Figure \ref{fig_many_hists}, while the histograms report the $M_{\star}$ and $L_{\rm [O~III]}$ (both parameters used during the match) and the SFR($H\alpha$) (same SFR used in Figure \ref{fig_radio_sample}), which is not used for the match.
    
    The upper-left plot shows that Sample~A has greater W$_{80}$ values at all annuli than non-AGN when only the $M_{\star}$ is considered during the match. We can see that the $L_{\rm [O~III]}$ is systematically lower for non-AGN. As mentioned above, \citet{Gatto_2024} find that their non-AGN control sample has [O~III] linewidths that correlate with [O~III] luminosity. If we include the $L_{\rm [O~III]}$ (bottom left plot of Figure \ref{fig_SampleA_vs_nonAGN}), the median W$_{80}$ value at all annuli of Sample A and the non-AGN sample behaves similarly. While there might be a small difference at small radii $R_{\rm eff} < 0.5$, at large radii, there is no difference between both samples. This suggests that the stacked W$_{80}$ radial profiles of optically selected AGN together with broad-line selected AGN can be easily reproduced by non-AGN hosts with the same distribution of mass $M_{\star}$ and $L_{\rm [O~III]}$.
    
    In contrast to Sample~A, Sample~B (see Figure \ref{fig_SampleB_vs_nonAGN}) shows higher W$_{80}$ values compared to non-AGN. Remarkably, the most significant difference between non-AGN and sample B is seen at the largest annuli, where high W$_{80}$ values are achieved by Sample B. Excluding optically (and broad-line) selected samples from our radio-selected AGN systematically removes AGN with low W$_{80}$ at larger annuli. For Sample~A, it appears that excluding radio-selected AGN from optically (or broad-line) selected samples consistently removes the W$_{80}$ kinematic excess at larger annuli.

\subsection{AGN versus AGN}
\label{sec:AGN_vs_AGN}

    In Figure \ref{fig_many_hists}, we find that our broad-line selected AGN tend to have larger $L_{\rm [O~III]}$ than optically-selected AGN galaxies. Therefore, in the top panel of Figure \ref{fig:agn_vs_agn}, we match the two samples within $39.4<$~log($L_{\rm [O~III]}$)~$<42.1$ and $10.4<$~log($M_{\star}$)~$<11.7$ (limits within which the parameter distribution can be matched) and compare the two samples: optically-selected AGN excluding broad-line selected AGN and vice-versa. We find that the median W$_{80}$ radial profile of broad-line-only AGN is similar to the optically-selected AGN (excluding broad-line AGN) when both samples are matched in host galaxy properties, with a small excess in the center for the broad-line selected AGN (see Figure \ref{fig:agn_vs_agn}). \citet{Gatto_2024} arrive at similar conclusions for their broad-line and optically-selected AGN, reporting no difference when comparing their $W_{80}$ distributions. The results remain unchanged if we also control for inclination ($b/a$), ruling out possible orientation effects. Therefore, we combine these samples when setting up Sample A and Sample B (in Section \ref{sec:AGN_vs_non-AGN}). 
    
    We now proceed with comparing Sample A with Sample B directly. As before, we also match the samples in $M_{\star}$ and $L_{\rm [O~III]}$, redshift and morphology. Note that, when excluding radio-selected sources from the optically-selected AGN catalog (Sample A), we do not claim that the remaining optically-selected AGN have no AGN-related radio emission, but we rather aim to investigate kinematic properties of a sample that would not have been detected as AGN through radio techniques and vice versa.

    In the bottom-left panel of Fig. \ref{fig:agn_vs_agn}, we present the W$_{80}$ radial profiles of Sample A and Sample B. Sample A is forced to match Sample B within $39.2<$~log($L_{\rm [O~III]}$)~$<41.5$ and $10.1<$~log($M_{\star}$)~$<11.7$. Sample B (represented in blue) shows elevated W$_{80}$ values across all annuli, notably at large $R_{\rm eff}$, aligning with the findings in Section \ref{sec:AGN_vs_non-AGN}. This comparison suggests that while optical and broad-line selection methods can identify AGN hosts with perturbed kinematics extending to large galacto-centric distances, the absence (or exclusion) of radio-selected AGN (as in Sample A, shown in gray) results in a population characterized by systematically reduced kinematic disturbances at these distances. Conversely, when optically and broad-line selected AGN are removed from a radio-selected sample (as in Sample B), the remaining AGN hosts predominantly exhibit significant kinematic perturbations, especially at extended $R_{\rm eff}$ scales. This suggests that AGN radio-selection techniques are sensitive to finding AGN hosts with disturbed kinematics over larger galacto-centric distances.

        A key takeaway message is that the selection technique is sensitive to the kinematics found in AGN galaxies and might also be sensitive to the evolutionary stage of AGN (see the Discussion section). We point out that employing alternative cutoff lines in our radio selection technique (e.g. 1.5~dex; see Section \ref{sec_radio_selection} and  Figure \ref{fig_radio_sample}) produces similar outcomes. As illustrated in Figure \ref{fig_many_hists}, a larger cutoff would also result in a cut on stellar mass. However, this also would substantially reduce the number of targets. Our results concerning radio AGN are very similar if we consider a different SFR estimator (as mentioned in Section \ref{sec_radio_selection}) when selecting our radio sample.

\subsection{SF galaxies vs AGN}

    We perform a similar comparison between AGN and (see Section \ref{optical_class}) SF galaxies (as classified by BPT diagnostics). We find that SF galaxies have a lower median W$_{80}$ radial profile compared to any of the AGN-selected samples (AGN were excluded from the SF galaxy sample, although only 15 AGN overlap with it). Furthermore, SF galaxies have indeed higher SFRs than our selected AGN and even higher SFRs when controlling for $M_{\star}$ and $L_{\rm [O~III]}$ to a specific AGN selected population (they also have younger D4000 ages, and higher H$\alpha$ equivalent widths). This suggests that SF galaxies (at least, BPT-classified) in MaNGA do not seem responsible for driving significant ionized gas outflow signatures, even when having significantly higher SFRs.

    Lastly, in Section \ref{sec_quality_cirteria}, we described that we only use galaxies with at least two available annuli where at least 10\% of their spaxels have S/N~$>7$. 
    This quality criteria results in some galaxies having no W$_{80}$ values in some of their annuli. To control for a possible impact of this effect, we study the same comparisons described in this Section with two samples: one where we use all the galaxies and all their spaxels that have a S/N~$>3$ (low S/N), and the other one where we only use galaxies that have at least six available annuli where at least 10\% of their spaxels have S/N~$>7$ (a high S/N constraint). Using the latter samples, we confirm that the behavior described in this Section is still present for all the comparisons.

\section{Discussion}
\label{sec_discussion}

\subsection{AGN selection and their integrated host galaxy properties}

    In this paper, we find that different AGN selection techniques select AGN samples that hardly overlap in more than 50\% of their targets. Similar results have been found in \citet{Oh2022} at $z<0.2$ when comparing X-ray selected to optically-selected AGN. Additionally, for higher redshifts ($ 0.25 <z< 0.8$), in a sample of mid-infrared, radio, and X-ray selected AGN, \citet{Hickox_2009} find AGN candidates hardly overlapping (their radio selection is at L$_{1.4~\text{GHz}}>10^{23.8}$~W~Hz$^{-1}$). These findings impose a clear challenge in AGN studies, since AGN found by the different selection techniques do not always trace the same host galaxy properties and / or AGN accretion state \citep{Hickox_2018}.

    We find that our radio-selected AGN are typically found below the main sequence of SF galaxies (SFMS). Similar results have been found specifically for MaNGA \citep{Comerford2020, Mulcahey_2022}, and other low redshift studies \citep[e.g.][]{Smolvic2009}. Accordingly,  \cite{Sanchez_2022} find that optically-selected AGN lie below the SFMS in the Green Valley. Additionally, \citet{Schawinski_2007} find that SF galaxies, composite and AGN (all optically selected) seem to follow an evolutionary sequence in the star formation and stellar mass plane, with SF galaxies having bluer colors and AGN found more in the transition zone. Our composite-selected targets are also found between SF and AGN-selected galaxies in the $M_{\star}$ versus SFR plane. \citet{Hickox_2009} find similar results, also in agreement with our mid-infrared selected targets, which they find to be more likely found in slightly more star-forming hosts. They propose an interpretation suggesting an evolutionary picture where, as star formation decreases, AGN accretion changes from optical or infrared-bright to optically faint radio sources. These findings suggest that AGN selection techniques are sensitive not only to the physical processes powering them but also to the stage of their duty cycle. We discuss this further in Section \ref{radio_discussion}.

\subsection{Spatially resolved ionized gas kinematics}
   
    We first investigated the radial properties of the [OIII] ionized gas kinematics of unmatched AGN and non-AGN samples, showcasing a diverse range of ionized gas kinematics (this was done before controlling for host galaxy properties; see Figure \ref{fig_W80s}). Non-AGN and SF galaxies exhibit less disturbed kinematics compared to all AGN samples (lower W$_{80}$ radial profiles). 
    
    When comparing AGN samples matched in $M_{\star}$ and L$_{\text{[O~III]}}$, intrinsic distinct kinematic behaviors emerge. Specifically, the exclusion of radio-selected AGN from an optical and broad-line selected AGN sample (Sample A) results in lower $\rm W_{80}$ values at greater galacto-centric distances, suggesting that much of the kinematic disturbances within an optically-selected sample are linked to the radio emission in AGN (see more discussion on the connection of outflows and radio emission in the next Section). 
    
    The analysis of Sample A also reveals that there is a population of non-AGN galaxies that can easily produce AGN-like W$_{80}$ profiles when controlling for host galaxy properties (see bottom-left panel of Figure \ref{fig_SampleA_vs_nonAGN}). Simulations suggest that kiloparsec-scale AGN-driven outflows can outlast the AGN activity phase, extending from a few to several orders of magnitude longer in duration \citep[a few Myr,][]{King_2011, Zubovas_2018}. For example, \citet{Zubovas_2022} predicts that fossil outflows (outflows living after the AGN switches off) could actually be more common than finding an outflow and an AGN in a galaxy simultaneously. Consequently, MaNGA non-AGN galaxies may include some galaxies showing fossil outflows. The possible presence of fossil outflows in MaNGA galaxies will be discussed in a future paper.

\subsection{Radio-selected AGN as tracers of the final phases of AGN evolution}
\label{radio_discussion}

    We find that AGN identified through radio techniques alone (Sample B) show notably stronger kinematics at larger R$_{eff}$ than any other AGN sample. The presence of AGN-related radio emission in AGN may, therefore, seem to trace AGN with more spatially extended outflows.
    
    One explanation for this behavior may be that sources with AGN-related radio emission trace host galaxies that have been experiencing AGN activity (or activities; see below) for a longer time. However, kinematic perturbances up to kpc scales would typically imply an active (AGN) phase longer than the duration of a typical AGN duty cycle. For example, \citet{King_2011} uses analytical models to study the outflow propagation during an AGN event. They show that an outflow with an initial velocity of a couple of hundred km~s$^{-1}$ in an AGN episode lasting about $\sim$1~Myr can last up to ten times more than the AGN itself, reaching several kpc.
     
     Alternatively, radio-selected AGN may be sensitive to AGN that have gone through multiple cycles of AGN activity.  Indeed, some galaxies show evidence of past and recurring AGN events 
     \citep[e.g.,][]{Schawinski_2015, Shulevski_2015, Vaishnav_2023}. Recent studies have used low-frequency (MHz) radio spectra combined with high-frequency spectra (GHz) to trace back emissions from previous activities \citep[e.g.,][]{Jurlin_2020}. A younger AGN phase is characterized by a peaked spectrum in the center, while a remnant from past events displays a more spread-diffuse emission. Therefore, if a combination of the latter is observed in one target, it can suggest the target is a strong candidate for a restarted AGN phase. In this context, \citet{Kukreti_2023} find that around 6\% of the targets in their sample (at  $10^{23}\text{W}~\text{Hz}^{-1}>\text{L}{1.4~\text{GHz}}>10^{26}\text{W}~\text{Hz}^{-1}$, $0.02<z<0.23$) are peaked sources classified as compact in GHz frequencies but have extended emission at MHz frequencies, suggesting them as restarted AGN candidates.

     Also, the simulations from \citet{Zubovas_2023} show that fossil outflows in gas-poor systems tend to last longer than in gas-rich hosts. Radio-selected AGN, indeed, are preferentially found in gas-poor galaxies. 
     
     With respect to an AGN's impact on its host galaxy, a recent review by \citet{Harrison_2023} discusses that simulations predict that feedback that leads to galaxy quenching does not come from a single AGN event but is rather a cumulative effect of multiple AGN episodes \citep[see also][]{Piotrowska_2022}. Thus, given the findings in our analysis, AGN selected through radio observations may preferentially trace galaxies that have experienced episodic AGN events \citep{Morganti_2017_2}. Sample B indeed contains host galaxies with older stellar populations (compared to Sample A), as traced by average D4000 measurements from the PiPE3D catalog.
     
     In a sample of radio-selected AGN from MaNGA MPL-8, \citet{Comerford2020} find that radio-mode AGN host galaxies reside preferentially in elliptical galaxies have more negative stellar age gradients with galacto-centric distance. The authors suggest that radio-mode AGNs may represent a final phase in the evolution of AGN.  In addition, \citet{Hickox_2009} propose a scenario where radio-AGN are key to the late stages of galaxy evolution, with them being, in general, more passive and low Eddington ratios than their infrared and optical counterparts. 
     In fact, radio-selected AGN typically have larger black hole masses \citep{Best_2005, Hickox_2009}. Interestingly, the latter parameter is found to be a strong predictor for galaxy quenching \citep{Piotrowska_2022}. These results are in line with the here presented work on the spatially resolved ionized gas kinematics in radio-selected AGN, which suggest that radio selection methods may be used to identify AGNs at a more advanced stage of their activity (and feedback) cycle. Lastly, we note that most of the removed radio-selected AGN (see Section \ref{sec_quality_cirteria} and Section \ref{sec_host_gal_properties}) are massive galaxies with low SFRs (see Figure \ref{fig_radio_sample}), located near the red sequence, suggesting an even later evolutionary phase.

    \subsection{The connection between radio-emission and outflow activity in AGN}
    
    Our results discussed above raise the question of what mechanisms are responsible for the observed radio emissions. The possible origins of radio emission in low-luminosity radio AGN is reviewed in \citet{Panessa_2019}. The review discusses several mechanisms such as by jets, winds, accretion disk corona, and star formation. For the context of our work, winds are discussed as a mechanism in which a shock is driven by the wind \citep[e.g.,][]{Riffel_2021} and produce radio emission due to the acceleration of relativistic electrons on sub-kpc scales. Similarly, in a small sample of AGN ($z<0.07$), \citet{Mizumoto_2024} found that NLR-scale shocks \citep[traced by {[Fe~II]/[P~II]}; see][]{Oliva_2001} are likely triggered by ionized outflows (traced by [S~III] in Mizumoto et al.).  
    
    Notably, in a sample of galaxies at $z<0.8$, \citet{Zakamska_2014} show that the radio luminosity in formally radio-quiet AGN correlates with the [O~III] velocity width, consistent with our findings. \citet{Zakamska_2014} propose two scenarios: one where radio emission is produced by accelerated particles as a result of shock fronts due to outflows (extended and diffuse radio emission), and another one where an unresolved radio jet (unresolved in FIRST/NVSS data) is launching an outflow (expected to be compact). We argue that both scenarios could simultaneously be present in one system (e.g., if the galaxy had more than one recent AGN event). High spatial resolution radio observations would be needed to distinguish between them.
    
    \citet{Rivera_2023} arrives at a similar conclusions by analyzing the C${\text{IV}}$ and [O~III] velocities (in a sample of $\sim$100 AGN). They discover minimal or no correlation between the C${\text{IV}}$ velocities and radio luminosity, in contrast to a connection between the [OIII] velocity width and radio luminosity. Given that C${\text{IV}}$ emission originate from within the broad-line region (sub-pc scales), and [O~III] emission traces ionized gas on galactic / kpc scales, \citet{Rivera_2023} conclude that the interplay between winds and radio luminosity predominantly occurs on these circumnuclear scales. Similarly, \citet{Liao_2024} not only shows that [O~III] velocity widths of AGN (in their sample: $z<1.0$, and a median log(L$_{[\text{O~III}]})\sim42.1$) correlate with radio emission but also that the conversion efficiencies align with those needed to account for the observed radio luminosities in galaxies exhibiting large [O~III] velocity widths. Their results also support the idea that AGN-driven outflows contribute to the radio emission in AGN.

    While the results discussed above suggest a connection between the radio emission and the ionized gas kinematics in AGN, we note that they were done predominantly using single fiber spectra and investigating higher redshift galaxies ($z\lesssim0.8$), averaging the gas kinematics over larger areas. Our work adds to the picture using a spatially resolved kinematic analysis and while we cannot exclude the presence of jetted radio-AGN in our radio-selected sample, our results also suggest that there is a strong connection between radio activity and ionized gas outflows in AGN.

\subsection{SF galaxies}

    We find that SF galaxies show less enhanced kinematic profiles when compared to AGN candidates, even when controlling for $M_{\star}$, L$_{[\text{O~III}]}$, morphology and redshift. A detailed comparison between Sample A and SF galaxies highlights that Sample A demonstrates significantly higher W$_{80}$ values within its central regions. In contrast, such differences fade at larger effective radii (R$_{eff}$), where the kinematic behaviors of both populations (Sample~A and the matched star-formation sample) align closely. Moreover, the matched SF galaxy sample exhibits higher star formation rates than Sample A (and higher than the SF sample before matching). In a larger sample ($>50000$) of local ($0.05<z<0.1$) SF galaxies, \citet{Yu_2022} studied the ionized gas kinematics of these galaxies, finding that they can indeed present outflow signatures. But the authors also show that the star-forming sample hardly ever reaches $\sigma>150$~km~s$^{-1}$, i.e., W$_{80} > 375$~km~s$^{-1}$. This is consistent with our findings. Therefore, we infer that the enhanced W$_{80}$ values in our AGN-selected population are likely driven by AGN and do not expect star-forming processes to play a significant role.

    However, the most massive ($M_{\star}>10^{11}M_{\odot}$) SF galaxies in our sample reveal remarkably high W$_{80}$ values (although not as high as AGN). \citet{Sabater_2019} found that 100\% of the galaxies with masses above this limit ($10^{11}M_{\odot}$) host radio AGN even though sometimes with radio luminosities (L$_{150~ \text{MHz}}>10^{21.5}$~W~Hz$^{-1}$, or L$_{1.4~\text{GHz}}\gtrsim10^{21}$~W~Hz$^{-1}$; most of our galaxies are above this limit). Indeed, $>50$\% of massive SF galaxies in our sample have radio detections, while only $\sim10$\% of lower mass SF galaxies ($< 10^{11}M_{\odot}$) have radio detections. This suggests that some of our massive SF galaxy populations may be AGN as well.

\section{Summary and conclusions}
\label{sec_summary}

    We have assembled a multi-wavelength AGN-selected sample for the SDSS-IV MaNGA-DR17, comprising 594 unique AGN identified through optical, hard X-ray, radio, infrared, and broad-line selection techniques. We seek to explore the extent to which ionized gas kinematics, as quantified by W$_{80}$ of [O~III]$\lambda$5007, is influenced by the diversity in AGN selection methods, thereby offering insights into feedback processes and the duty cycle of AGN activity. To do so, we fit up to two Gaussian components to the [O~III]$\lambda$5007 emission line region in all spaxels ($S/N>7$; see Section \ref{sec_quality_cirteria}) of each galaxy and derive the W$_{80}$ velocity widths (see Section \ref{sec_W80_differences}). We then map the spatial distribution of this parameter for each galaxy. Furthermore, we create W$_{80}$ radial profiles and stack them according to each defined AGN subsample. Our findings are summarized as follows:

\begin{itemize}
    \item We find that different AGN selection techniques do not completely overlap with each other. Overlap ranges from $\sim$34\% (e.g., between radio and optical selection) up to $\sim$80\% (the latter percentage only achieved by the X-ray selection, although it is the smallest sample).

    \item The different AGN populations are found in galaxies with different host galaxy properties. The most significant differences are found in the distribution of L$_{[\text{O~III}]}$, EW($H\alpha$), D4000, and the $W_{80}$ radial profiles.

    \item Regarding AGN vs. non-AGN: Regardless of the selection technique, all AGN populations show more perturbed ionized gas kinematics (traced by $W_{80}$) at all annuli when compared to non-AGN of similar $M_{\star}$ of the host, redshift, and morphology. These kinematic differences become less pronounced when L$_{[\text{O~III}]}$ is taken into account in the non-AGN control sample. Remarkably, the differences between AGN and non-AGN disappear when we compare pure optical (BPT and broad-line, but exclude radio-detect AGN: Sample~A) AGN to non-AGN (see Section \ref{sec:AGN_vs_non-AGN}). We suggest that some non-AGN may host fossil outflows (i.e., relic outflows of a past AGN phase), which may outnumber outflows in currently active AGN \citep{Zubovas_2022}.

    \item Regarding AGN vs AGN: Our different AGN samples display not only hosts with different properties but also hosts with differences in the stacked radial profiles of their kinematic signatures. Interestingly, when controlling for host galaxy properties, we find that removing radio-selected-AGN from optically selected candidates leaves a sample (Sample A) of galaxies that lack significantly high $W_{80}$ at high R$_{eff}$, suggesting  that much of the kinematic disturbances within an optically-selected sample are linked to the radio emission in AGN. In addition, radio-selected AGN show more enhanced ionized gas kinematics at all radii and their hosts show evidence of older stellar populations. Our results support a scenario in which radio selection methods may be used to identify AGNs at a more advanced stage of their activity (and feedback) cycle.

    \item AGN vs star-forming: SF galaxies in our sample do not show significant kinematic signatures in the ionized gas compared to AGN (regardless of the selection technique; see Section \ref{sec_2}).
    We highlight that when controlling for L$_{[\text{O~III}]}$ and $M_{\star}$ when comparing AGN to non-AGN, SF galaxies tend to have significantly larger SFR(H$\alpha$)) than AGN. We conclude that in our sample, the main driver of the enhanced kinematic signatures in AGN cannot be accounted for by star formation processes alone.

\end{itemize}

    Our study shows that a given AGN selection technique can impact what sort of ionized kinematic signatures are found in their host galaxies. Our results are tested in low-redshift ($z <0.1$) galaxies with low- to intermediate luminosities. The impact of AGN selection techniques could be more significant at higher redshift. Moreover, our results highlight the importance and utility of spatially resolved spectroscopy.

\begin{acknowledgements}

      D.W. acknowledges support through an Emmy Noether Grant of the German Research Foundation, a stipend by the Daimler and Benz Foundation and a Verbundforschung grant by the German Space Agency.\\

      M.A. extends gratitude to the GALENA research group for their invaluable discussions, which have significantly shaped the ideas presented in this paper.\\ 

      J.M.C. is supported by NSF AST-1714503 and NSF AST- 1847938.\\

      RAR acknowledges the support from Conselho Nacional de Desenvolvimento Cient\'ifico e Tecnol\'ogico (CNPq; Proj. 303450/2022-3, 403398/2023-1, \& 441722/2023-7), Funda\c c\~ao de Amparo \`a pesquisa do Estado do Rio Grande do Sul (FAPERGS; Proj. 21/2551-0002018-0), and CAPES (Proj. 88887.894973/2023-00).\\

      This project makes use of the MaNGA-Pipe3D dataproducts. We thank the IA-UNAM MaNGA team for creating this catalogue, and the Conacyt Project CB-285080 for supporting them.\\

      Funding for the Sloan Digital Sky Survey IV has been provided by the Alfred P. Sloan Foundation, the U.S. Department of Energy Office of Science, and the Participating Institutions. SDSS-IV acknowledges support and resources from the Center for HighPerformance Computing at the University of Utah. The SDSS web site is www.sdss.org.\\

      SDSS-IV is managed by the Astrophysical Research Consortium for the Participating Institutions of the SDSS Collaboration including the Brazilian Participation Group, the Carnegie Institution for Science, Carnegie Mellon University, the Chilean Participation Group, the French Participation Group, Harvard-Smithsonian Center for Astrophysics, Instituto de Astrof\'isica de Canarias, The Johns Hopkins University, Kavli Institute for the Physics and Mathematics of the Universe (IPMU) / University of Tokyo, the Korean Participation Group, Lawrence Berkeley National Laboratory, Leibniz Institut f\"ur Astrophysik Potsdam (AIP), Max-Planck-Institut f\"ur Astronomie (MPIA Heidelberg), Max-Planck-Institut f\"ur Astrophysik (MPA Garching), Max-Planck-Institut f\"ur Extraterrestrische Physik (MPE), National Astronomical Observatories of China, New Mexico State University, New York University, University of Notre Dame, Observatario Nacional / MCTI, The Ohio State University, Pennsylvania State University, Shanghai Astronomical Observatory, United Kingdom Participation Group, Universidad Nacional Autonoma de M\'exico, University of Arizona, University of Colorado Boulder, University of Oxford, University of Portsmouth, University of Utah, University of Virginia, University
\end{acknowledgements}

\bibliographystyle{aa}
\bibliography{bibliography}

\begin{appendix}

\section{Targets in our sample}
\label{appendix_repeated_targets}

    Due to the amount of data, and catalogs available for MaNGA, this Section details our sample. We intend to remove duplicate observations to be specific about which targets we use since we require specific host-galaxy properties from the galaxies.

    \begin{table}[htbp]
    \centering
    \caption{Plate-IFU pairs of repeated observations.}
    \begin{tabular}{l|l}
    \hline\hline
    10513-1901 --- 9512-6104 & 7963-12702 --- 8651-12702 \\
    10513-3702 --- 9512-6103 & 7963-12704 --- 8651-12704 \\
    10843-12704 --- 11866-9101 & 7963-12705 --- 8651-12705 \\
    10843-6103 --- 11866-1901 & 7963-3701 --- 8651-3701 \\
    11016-12705 --- 11827-9101 & 7963-3704 --- 8651-3704 \\
    11016-1901 --- 8309-6101 & 7963-6101 --- 8651-9102 \\
    11016-1902 --- 8309-1902 & 7963-6102 --- 8651-1902 \\
    11016-3702 --- 8309-12705 & 7963-6103 --- 8651-6104 \\
    11016-6101 --- 11827-6103 & 7963-6104 --- 7964-12705 \\
    11016-6104 --- 11827-6104 & 7963-9101 --- 8651-6102 \\
    11017-12703 --- 11758-3702 & 7963-9102 --- 8651-6103 \\
    11017-1902 --- 8319-6104 & 7964-12701 --- 8651-12703 \\
    11017-9101 --- 11758-3701 & 7964-3702 --- 8651-3703 \\
    11757-1902 --- 11868-12705 & 8239-6104 --- 8567-12702 \\
    11823-3703 --- 11950-1902 & 8247-3702 --- 8249-3701 \\
    11823-6104 --- 11950-1901 & 8249-12705 --- 8250-3702 \\
    11823-9102 --- 11950-3701 & 8249-6104 --- 8250-9101 \\
    11827-12701 --- 8325-12704 & 8256-12701 --- 8274-12701 \\
    11827-3701 --- 8325-6103 & 8256-12702 --- 8274-12702 \\
    11827-3703 --- 9864-3701 & 8256-1901 --- 8274-1901 \\
    11827-3704 --- 8326-1901 & 8256-3701 --- 8274-3701 \\
    11827-9102 --- 8326-9101 & 8256-3702 --- 8274-3702 \\
    11838-12703 --- 11865-1901 & 8256-9102 --- 8274-9102 \\
    11838-3703 --- 11865-9101 & 8261-1901 --- 8262-1901 \\
    11867-12702 --- 12511-1902 & 8309-12702 --- 9884-1902 \\
    11867-12703 --- 12511-3703 & 8312-12703 --- 8550-9102 \\
    11867-6101 --- 9512-1901 & 8319-1902 --- 8324-1901 \\
    11867-6103 --- 9512-3702 & 8319-3704 --- 8324-9102 \\
    11867-6104 --- 9512-3704 & 8325-12703 --- 8326-12701 \\
    11867-9101 --- 9512-12701 & 8325-3704 --- 8328-1901 \\
    11867-9102 --- 9512-3701 & 8326-3701 --- 8329-1901 \\
    11940-6104 --- 12667-3701 & 8328-3704 --- 8329-12702 \\
    11946-6101 --- 12667-1902 & 8329-3701 --- 8333-12701 \\
    11947-3702 --- 12675-1902 & 8329-3703 --- 8333-12704 \\
    11948-12703 --- 12675-3704 & 8329-3704 --- 8333-3702 \\
    11949-1902 --- 8613-1901 & 8454-6103 --- 8456-6104 \\
    11949-3703 --- 8613-6103 & 8459-3701 --- 8461-6104 \\
    11978-6101 --- 9894-3702 & 8459-3702 --- 8461-3704 \\
    11978-6104 --- 9894-1901 & 8459-3704 --- 8461-12703 \\
    12066-1901 --- 8652-3702 & 8484-9101 --- 8555-3704 \\
    12066-3704 --- 8652-12701 & 8555-12701 --- 8600-9102 \\
    12667-3704 --- 12675-3702 & 8588-3701 --- 8603-12701 \\
    7815-12701 --- 8618-1902 & 8596-12701 --- 8598-12703 \\
    7815-12702 --- 7972-12705 & 8596-12702 --- 8598-9102 \\
    7815-12705 --- 8618-6101 & 8600-1902 --- 8979-3703 \\
    7815-1902 --- 8618-6103 & 8600-3702 --- 8979-12704 \\
    7815-6101 --- 7972-3701 & 8606-6104 --- 8614-3702 \\
    7815-9101 --- 7972-12704 & 8651-6101 --- 9191-3703 \\
    7958-1901 --- 9185-1901 & 8950-12702 --- 8951-12704 \\
    7958-3703 --- 9185-3702 & 8996-12705 --- 8997-12701 \\
    7960-12702 --- 9185-3704 & 8998-3703 --- 8999-9101 \\
    7962-3701 --- 9085-3701 & 9031-12701 --- 9036-12703 \\
    7962-6101 --- 9085-3703 & 9031-12705 --- 9036-6101 \\
    7962-6104 --- 9085-3704 & 9031-3701 --- 9036-1901 \\
    7963-12701 --- 8651-12701 & 9031-3704 --- 9036-3703 \\
    \hline
    \end{tabular}
    \label{tab:repeated_observations}
    \end{table}

    We start defining the sample with all the targets present on \citet{Sanchez_2022} value-added catalog, which starts with a sample of 10~220 galaxies. We use the plate-ifu as the main identifier of our targets, given that MaNGA has a number of repeated observations (some with the same MaNGAID). We follow MaNGA's steps to mask the sample for unique galaxies\footnote{\url{https://www.sdss4.org/dr17/manga/manga-tutorials/drpall/}}, reducing the sample to 9~995 targets. The sample gets reduced to 9~992 galaxies because three of them had no data stored on the public website of the data reduction pipeline: 11939-1901, 11949-1901, and 8626-9102 (also reported in a list of targets that failed to be analyzed by the DAP\footnote{\url{https://www.sdss4.org/dr17/manga/manga-caveats/}}.
    
    In \citet{Sanchez_2022}, a table showing duplicate observations is reported, and similarly, a table of duplicates is also reported on the latter website (warning about duplicate galaxies with different MaNGA IDs). We note that some targets present in the MaNGA's duplicate table are not present in Sanchez's table. Therefore, we merge both repeated target tables and select from each pair the plate-ifu of which had more available annuli with higher S/N when measuring their W${80}$. This removes twenty more galaxies, leaving our sample with 9~972. We further double-checked for duplicate observations matching targets by MaNGA-ID and ensuring that the coordinates were consistent with each other and found more repetitions in the sample. We show these duplicate observations in Table \ref{tab:repeated_observations}, while some observations are repeated more than two times, as shown in Table \ref{tab:repeated_observations_long} (most of the targets in both tables come from cluster ancillary programs discussed in the drpall website mentioned above). As before, we remove these, keeping the one that offers a better quality of W$_{80}$. With the latter, we end up with 9~853 targets. Finally, we remove targets flagged by the MANGA\_DRP3QUAL as CRITICAL by the DRP. This leaves us with a final sample of 9777 galaxies.

\begin{table}[htbp]
\caption{Repeated observations with more than two elements.}
\centering
\begin{tabular}{l}
\hline
\hline
7963-3702 --- 8651-1901 --- 9191-3702 \\
8256-12703 --- 8274-12703 --- 8451-12704 \\ 
8256-12704 --- 8274-12704 --- 8451-12701 \\
8256-12705 --- 8274-12705 --- 8451-12702 \\ 
8256-1902 --- 8274-1902 --- 8451-1902 \\
8256-3703 --- 8274-3703 --- 8451-3703 \\ 
8256-3704 --- 8274-3704 --- 8451-3704 \\ 
8256-6101 --- 8274-6101 --- 8451-6101 \\ 
8256-6102 --- 8274-6102 --- 8451-3702 \\ 
8256-6103 --- 8274-6103 --- 8451-6102 \\
8256-6104 --- 8274-6104 --- 8451-6103 \\ 
8256-9101 --- 8274-9101 --- 8451-9101 \\\hline
8479-3703 --- 8480-3701 --- 8587-3702 \\
8953-3702 --- 9051-6103 \\\hline 

\end{tabular}
\label{tab:repeated_observations_long}
\tablefoot{the last row is a target repeated five times.}
\end{table}
   
    The quality criteria used in our analysis remove a number of extra galaxies from the study (see Section \ref{sec_quality_cirteria}). We do not analyze separately additional galaxies \citep[e.g., if more than one galaxy was found in a specific plate-ifu,][]{Pan_2019} found in the same IFU, and do not include any special treatment where this happens. 

\section{Fitting procedure details}

Our pipeline starts by subtracting the stellar continuum \citep[provided by the DAP][]{MangaDAP} from all spectra and moving each to its rest frame. We focus on the  4920-5080~\r{AA} region and subtract an additional continuum component from a 1D polynomial using two spectral windows (the first between 4870-4900~\r{AA} and the second between 5040-5100 \r{AA}). We execute the fitting two times: the first using a single Gaussian for each emission line and the second allowing two Gaussians for each emission line to account for possible asymmetries in the line profiles.

Below, we list the constraints used during the fitting procedure. The model with just one Gaussian profile has three free parameters to be fitted: amplitude, width, and systemic velocity, denoted by $A$, $\sigma$, and $\mu$, respectively. The details are given below:

\begin{itemize}
    \item The [O~III]~4959,5007~\r{AA} doublet is fixed to the theoretical flux ratio of 2.98 \citep[$\lambda$5007/$\lambda$4959;][]{Storey_2000,Laker_2022}.
    \item The velocity dispersion ($\sigma$) and systemic velocity ($\mu$) of both [O~III]~4959~\r{AA} and 5007~\r{AA} are tied to the same value, which will be a free parameter on the fitting procedure.
    \item We limit velocity values to $0<\sigma<1000$~km~s$^{-1}$, and $-1000<\mu<1000$~km~s$^{-1}$

\end{itemize}

In the two-Gaussian model, the fitting procedure has six free parameters. The first Gaussian component with $A$, $\sigma$, and $\mu$, and similarly, the second with $A_{w}$, $\sigma_{w}$, and $\mu_{w}$ (for the amplitude, width and offset of the ``wing'' component). We adopt the same considerations as listed above, and we add the following to the second Gaussian component:

\begin{itemize}
    \item The amplitude $A_{w}$ is a fraction of the main Gaussian amplitude ($A$), constrained between 0 and 1.
    \item The velocity dispersion $\sigma_{w}$ is forced to be higher and up to $1500$~km~s$^{-1}$ to avoid fitting noise.
    \item The systemic velocity  $\mu_{w}$ can be blue or redshifted up to $1000$~km~s$^{-1}$ from the main Gaussian's offset $\mu$.
\end{itemize}

A visual inspection of many of our results motivated us to add an extra condition to prevent the second component from fitting noise. To do this, we impose an additional condition to decide whether to use one or two Gaussians for the emission line. The second Gaussian component (after fitted) should have at least a $S/N_{[O~III]}$>3. If this $S/N$ requirement is not satisfied, the emission line is kept fitted with only one Gaussian.

We store each fitted parameter in maps (for the single- and double-Gaussian fitting procedures), including the reduced-chi-square provided by LMFIT. From the resulting maps, we construct the L$_{[O~III]}$ map (the sum of both components' fluxes in the case of the double-Gaussian model) and a non-parametric emission-line width map. To capture the emission-line width of a complex profile (e.g., a mixture of two Gaussian profiles) and reduce being influenced by the criteria of our fitting procedure, non-parametric measurements are routinely adopted \citep[e.g.,][]{Zakamska_2014, Wylezalek2020}. Specifically, we use the width that encloses the 80\% of the total flux, known as the non-parametric W$_{80}$ parameter \citep[see the details in][]{Liu_2013}. This parameter aims to prevent discarding information from the additional components of a profile composed of multiple components.

\begin{figure}
	\includegraphics[width=\columnwidth]{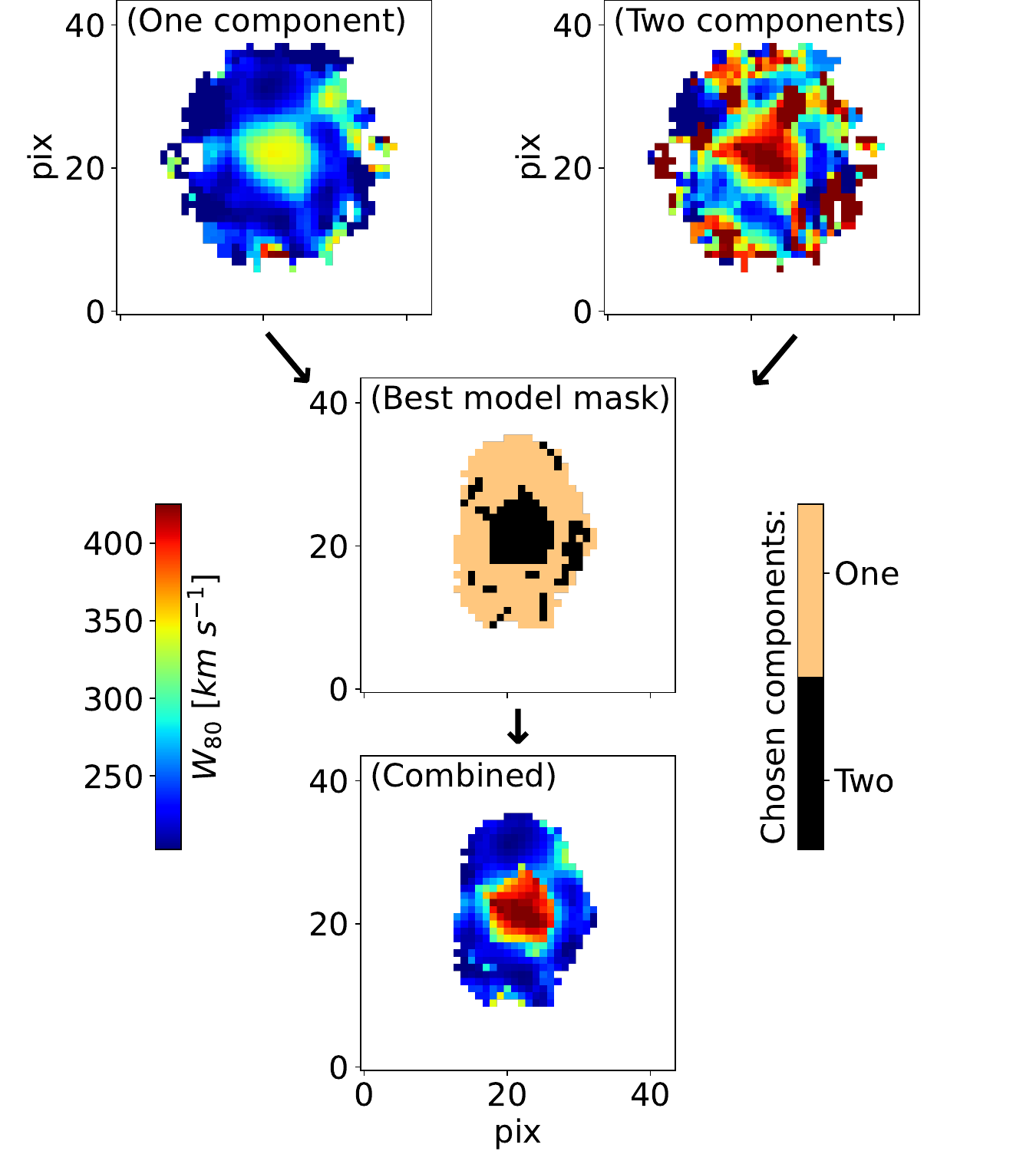}
 
    \caption{Output for MaNGA plate-IFU: 8244-3702. Final W$_{80}$ map (bottom plot) combined from the W$_{80}$ map of each model (top plots) based on the best $\chi^{2}_{red}$ mask (middle plot) and an additional S/N cut on the second Gaussian component (see Section \ref{sec_3.1}). All these Figure's W$_{80}$ maps have the same contrast colored following the same colorbar (on the middle-left)}.
    \label{fig_best-model}
\end{figure}

Finally, the decision to keep one fitting procedure from both models is based on the best-reduced chi-square \citep[the one closer to 1;][]{red_chi}. With the latter, we construct best-model mask maps (see Figure \ref{fig_best-model}), which are used to combine the results from the two fitting techniques into one containing the results of the models that fitted the spectral region the best (in the Figure, we show this for the [O~III]$\lambda$5007 W$_{80}$). The same best-model map creates the combined L$_{[O~III]}$ map for each galaxy. From these two maps, we extract the following parameters:

\begin{itemize}
    \item W$_{80}$ radial profiles for each map: average W$_{80}$ at elliptical ring apertures with a step of 0.25~R$_{eff}$ from the center of each target (see below).
    \item L$_{[O~III]}$ averaged at a radius of 0.5~R$_{eff}$.
\end{itemize}

\label{app_fitting_procedure}

\section{Binning parameters}
\label{app_binning}

\begin{figure*}
	\includegraphics[width=500pt]{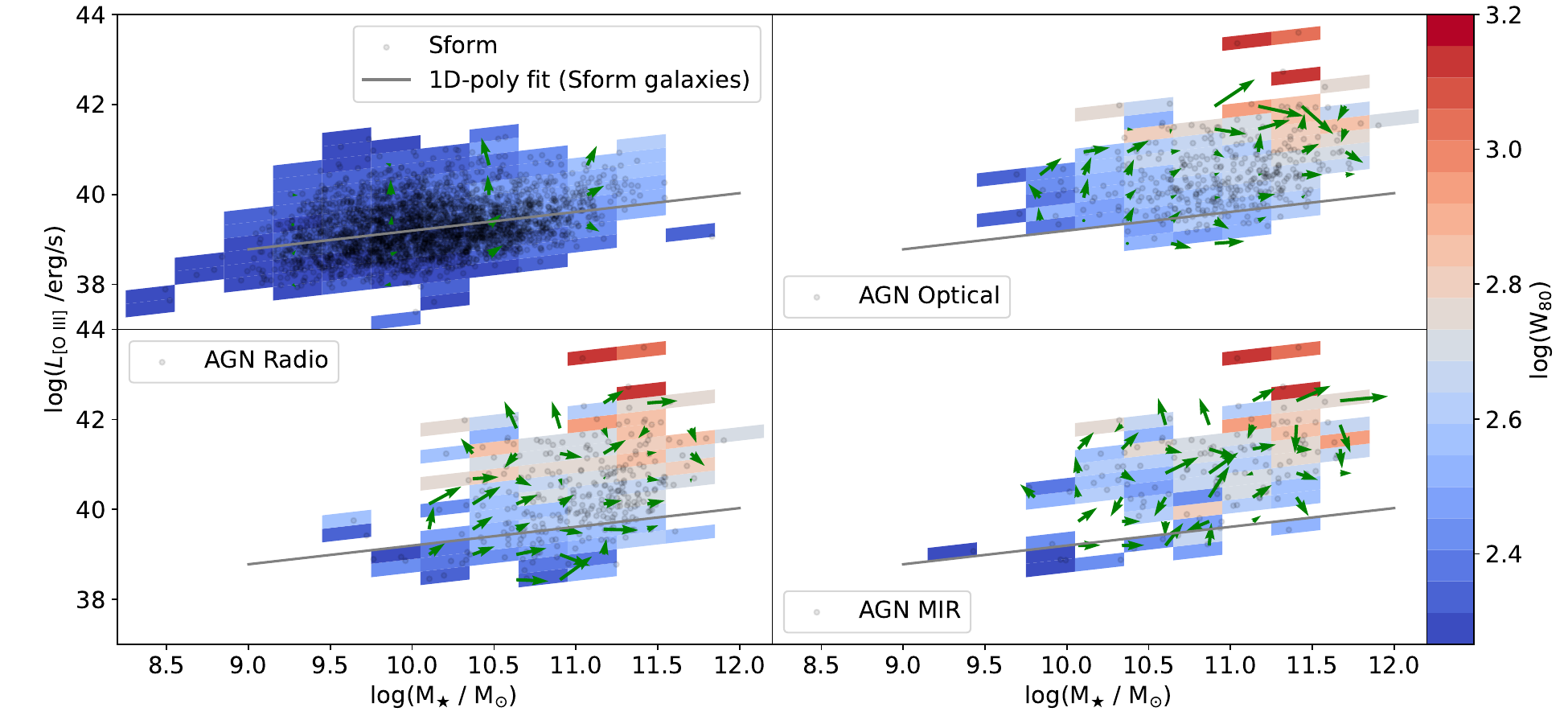}

    \caption{Average W$_{80}$ binned on a plane of $M_{\star}$ vs. L$_{[\text{O~III}]}$. The bins have a size of 0.3 dex in each parameter, colored by the strength of the W$_{80}$. The scatter dots show the distribution of a specific galaxy population and the line shows a 1D polynomial fitted to the location of the SF galaxies. The green arrows in the plots illustrate the gradient change of W$_{80}$ in the parameter space, with the arrowhead indicating the direction and the arrow's size representing the magnitude.}
    \label{fig_bins_2}
\end{figure*}

\begin{figure*}
	\includegraphics[width=500pt]{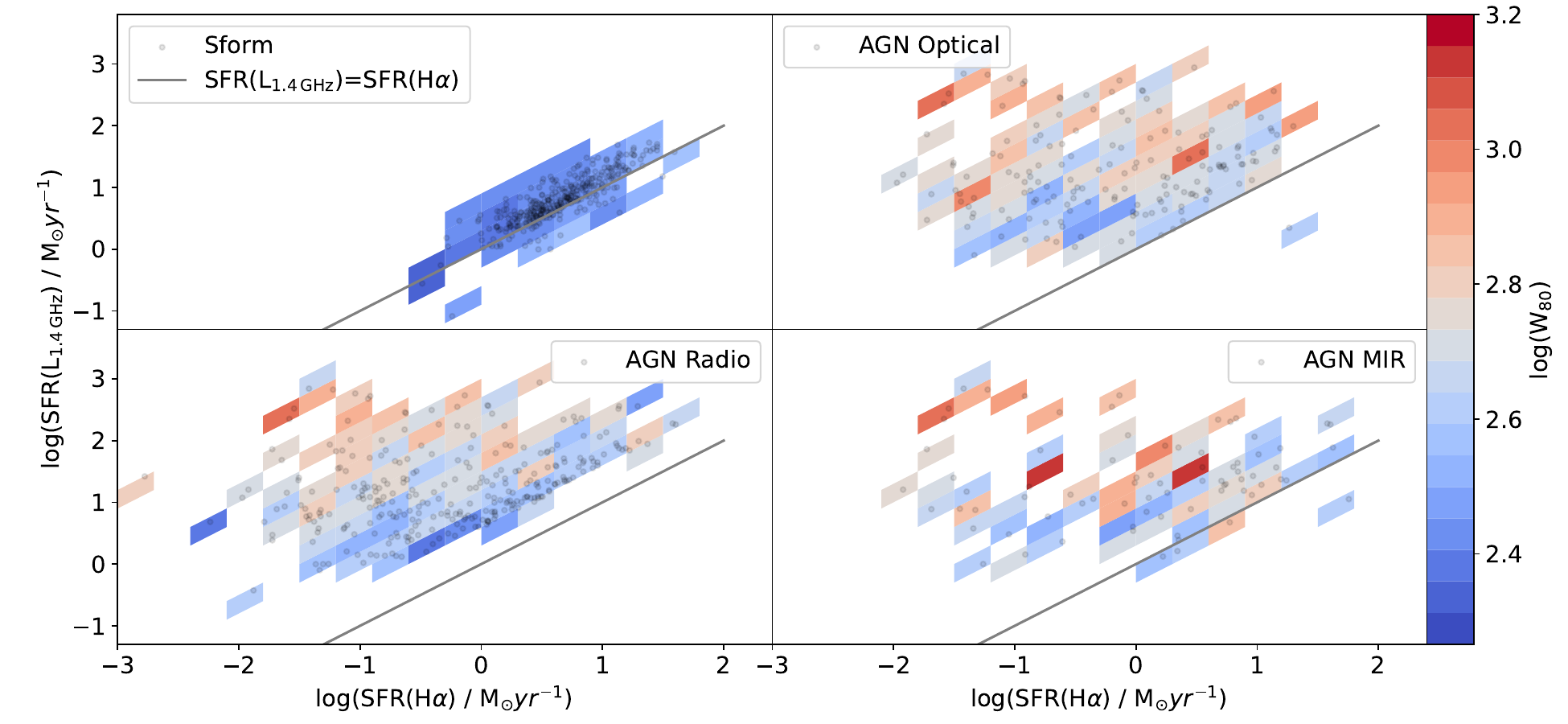}

    \caption{Average W$_{80}$ binned on a plane of star formation rate measured from L$_{rad}$ and L$_{H\alpha}$. The bins have a size of 0.3 dex in each parameter, colored by the strength of the W$_{80}$. The scatter dots show the distribution of a specific galaxy population and the line shows the 1-to-1 relation if both SFR tracers are equal.}
    \label{fig_bins_1}
\end{figure*}

We show how the ionized gas kinematics from the [O~III] emission line (traced by the central W$_{80}$ averaged over an aperture of 0.5~R$_{eff}$) changes (for different galaxy populations) when observed in different parameter spaces of host galaxy properties. This is shown in Figure A1.\ref{fig_bins_1} and Figure A1.\ref{fig_bins_2}.
 When looking at SF galaxies, in most cases, there is no significant evolution in their W$_{80}$. Conversely, in the case of AGN-selected galaxies, stronger W$_{80}$ values are found as we move to a specific direction of the parameter spaces. This is visually represented in Figure A1.\ref{fig_bins_2}, where minimal gradients are observed for SF galaxies, whereas AGN-selected galaxies show not only larger gradients but also a distinct trend towards increased stellar mass ($M_{\star}$) and L$_{[\text{O~III}]}$ luminosity.

\end{appendix}

\end{document}